\documentclass[12pt]{iopart}
\usepackage{graphicx}
\usepackage{iopams}

\begin{document}
\title[Nature of the ground state of 1$T$ metal dichalcogenides]{Topical review: the nature of the ground state and possibility of a quantum spin liquid in 1$T$ metal dichalcogenides}

\author{C J Butler \footnote{Author to whom any correspondence should be addressed}, M Naritsuka, T Hanaguri}

\address{RIKEN Center for Emergent Matter Science, 2-1 Hirosawa, Wako, Saitama 351-0198, Japan}
\ead{christopher.butler@riken.jp}
\ead{masahiro.naritsuka@riken.jp}
\ead{hanaguri@riken.jp}

\begin{abstract}

The compounds 1$T$-Ta$X_{2}$ ($X$ = S, Se) and 1$T$-NbSe$_{2}$ have been considered as potential hosts of a quantum spin liquid phase. This is based on the widely held view that the Mott-Hubbard mechanism drives the insulating behaviour of its charge density wave ground state, resulting in localized spins, interacting antiferromagnetically, on a geometrically frustrated lattice. However this assumes layer-wise independent behaviour. A growing body of evidence shows not only that inter-layer interactions are very significant in 1$T$-TaS$_{2}$, but also that they mediate some of its most interesting functional properties. Here we offer a perspective from the point of view of scanning tunnelling microscopy that helps to visualize the microscopic degrees of freedom of inter-layer interactions in bulk 1$T$-TaS$_{2}$, and the associated impact on the local density-of-states, including the occurrence of multiple distinct insulating phases. We consider to what extent the bulk of 1$T$-TaS$_{2}$, and its surface terminations can be considered as Mott insulating and whether, or where, quantum spin liquid behaviour might persist. To better understand the bulk behaviour we also draw insights from measurements on isolated monolayers of 1$T$-Ta$X_{2}$ and 1$T$-NbSe$_{2}$, where the confounding complications of inter-layer interactions are absent. We highlight some outstanding questions raised by a comprehensive evaluation of the experimental results, and finally suggest future experiments that could address them.

\end{abstract}
\vspace{2pc}

\submitto{\JPCM}
\maketitle

\section{Introduction}

The 1$T$ polytype of particular metal dichalcogenides have been identified as potential hosts of the elusive quantum spin liquid (QSL), a macroscopically entangled phase of electronic matter that may realize such unusual phenomena as emergent gauge fields, fractional quasiparticles \textit{via} spin-charge separation and non-abelian anyons, among others \cite{Balents2010,Savary2017,Zhou2017}. This idea stems from the recognition by Lee and Law that in the low-temperature charge density wave (CDW) phase of 1$T$-TaS$_{2}$, each layer realizes a Mott state on a geometrically frustrated triangular lattice \cite{Law2017}. A Mott state is often inferred for a system that has a half-filled band (i.e. with an odd number of electrons per unit-cell), but that is nonetheless observed to be insulating \cite{Mott1937}, and when considering only any single layer of 1$T$-TaS$_{2}$ this at first appears to be the case. However, this neglects the real likelihood that inter-layer interactions play a very important role in determining the electronic properties of bulk 1$T$-TaS$_{2}$, including determining what kind of insulator the bulk is. Indeed much of the research activity around 1$T$-TaS$_{2}$, of which the pursuit of a QSL is only a small part, focuses on the control of its electronic properties, using light, electric field or pressure, through manipulation of various microscopic configurations of the CDW distinguished in part through inter-layer degrees of freedom \cite{LiNaik2020,Wang2020,Hua2025}.

State-of-the-art measurement techniques are increasingly capable of characterizing the three-dimensional CDW configuration, and in the case of scanning tunnelling microscopy (STM), simultaneously probing the associated electronic density-of-states (DOS). STM measurements, as we will see below, can directly associate the inter-layer stacking configuration of the CDW at or near the surface with local transitions between metallic and insulating phases, and even between multiple insulating phases with different band gaps, distinct tunnelling spectra and, perhaps, qualitatively different insulating mechanisms \cite{Butler2020}. STM measurements are also increasingly used to probe the electronic structure of single- and few-layer samples of 1$T$-Ta$X_{2}$ and 1$T$-NbSe$_{2}$, from which insights can be drawn about the nature of the bulk 1$T$-TaS$_{2}$, but which also have their own interesting properties. Most notably, in single-layer films the reasons to anticipate a Mott state and QSL are revived.

This Topical Review is structured as follows: In Section 2 we will give an overview of the various relevant mechanisms that can drive insulating behaviour, and of the challenge of identifying which mechanism is in play in real materials. We will also describe the basics of spin liquids, including the prerequisite material properties such as Mott insulating behaviour, and some of the ways of detecting them. In Section 3 we will recount the development in understanding of the low-temperature CDW phase of 1$T$-TaS$_{2}$ and establish the necessary framework to understand the characterization and role of inter-layer effects in 1$T$-TaS$_{2}$. In Section 4 we review recent insights into inter-layer effects gained in large part from STM measurements, and we discuss the impact of the observed microscopic degrees of freedom for the nature of the insulating state. In Section 5 we will review the electronic properties of single- and few-layer 1$T$-Ta$X_{2}$ and 1$T$-NbSe$_{2}$, and in Section 6 we will discuss evidence for a QSL both in at the surface of the bulk and in monolayer samples. Here we also present some recent observations of super-modulations of the local DOS in monolayer 1$T$-TaSe$_{2}$, beyond the CDW. Finally, in Section 7 we will draw tentative conclusions, identify outstanding questions prompted by the observations reported so far, and make suggestions for future measurements.

\section{Background}

\subsection{Mechanisms of insulating behaviour}

Whether bulk 1$T$-TaS$_{2}$ can host a QSL, even in principle, depends on what kind of insulator it is. When insulating behaviour is observed in a given material, there are in fact many mechanisms that can be invoked in order to explain it, and multiple mechanisms can coexist in the same material, greatly obfuscating a full understanding. 
Thus there are a few materials for which, even after many years of investigation, the mechanism underpinning even their most basic electronic behaviour remains poorly understood \cite{Wentzcovitch1994,Weber2012,Brito2016}, and below it will be seen that 1$T$-TaS$_{2}$ is such a material.

\subsubsection{Band insulators}
The most basic explanation for why some materials are insulators and others are metals comes from Bloch band theory. This states that if a material has an odd number of electrons per unit-cell, one electronic band must be only half-filled, crossing the Fermi energy and allowing conduction. On the other hand, if there are an even number of electrons in each unit-cell it will tend to have a filled valence band, and the Fermi energy will be somewhere above this valence band and below an empty conduction band \cite{Hook_and_Hall_book}. Figures 1(a) and 1(b) illustrate these respective cases. 

\subsubsection{Peierls and charge density wave insulators}
Metals can exhibit a variety of instabilities that cause a restructuring of the unit-cell, and therefore of the Brillouin zone (BZ) and band structure. The textbook example is the Peierls instability, which occurs in a one-dimensional metallic chain due to the divergence of the Lindhard susceptibility $\chi(q)$ at a wavenumber $q = 2k_{\mathrm{F}}$ that fulfills the `nesting' condition of the one-dimensional Fermi surface \cite{Misra_book}. The diverging susceptibility renders the system unstable against the formation of a $q = 2k_{\mathrm{F}}$ modulation in the charge density $\rho_{\mathrm{e}}(q)$. Through electron-phonon (\textit{e-ph}) coupling, the ionic lattice responds with a periodic lattice distortion also at $2k_{\mathrm{F}}$ \cite{Kohn1959}. The result, illustrated in Fig. 1(c), is that the crystal's unit-cell undergoes a doubling in size, sometimes called Peierls dimerization, causing the BZ to halve in size. This enforces gaps where metallic bands of the un-distorted structure intersect the new zone boundary. From the point of view of Bloch band theory, the one-dimensional chain has undergone a transition from half-filled to filled by doubling the number of electrons per unit-cell. 

Many materials exhibit CDW order, especially the two- and quasi-two dimensional transition metal dichalcogenides. The Peierls mechanism is not fully adequate to explain the occurrence of CDWs in two dimensions, and even less so in three dimensions. Nevertheless, because CDWs often seem to be associated with nesting vectors supported by the materials' normal state Fermi surfaces, and with corresponding reductions of the phonon frequency, CDWs are often stated to have a `Peierls-like' underlying mechanism. However the full microscopic details and mechanisms underpinning most observed CDWs remain elusive even after intensive investigation \cite{Hwang2024}. Regardless of the driving mechanism, the outcome remains that upon formation of a CDW the BZ shrinks to a fraction of its former size and causes the band structure to be `folded' in such a way that new gaps, termed CDW gaps, are opened.

\begin{figure}
\centering
\includegraphics[scale=1]{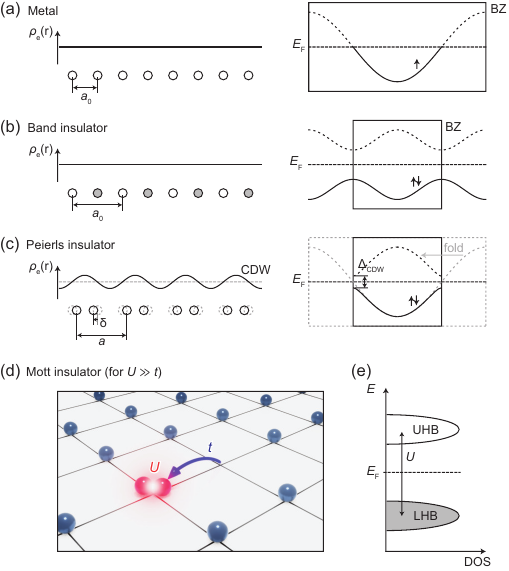}
\caption{\label{fig:1}
Mechanisms of insulating behaviour (a) A sketch of a one-dimensional metallic chain, its charge density $\rho_{\mathrm{e}}(r)$, and its band structure plotted in the first Brillouin zone (BZ), hosting a half-filled band. (b) An insulator with an even number of orbitals per unit-cell and the resulting filled valence band. (c) A one-dimensional metallic chain undergoing a Peierls transition. The doubling of the unit-cell size causes the shrinking of the BZ to half its former size, the folding of the band structure into the new BZ, the opening of a gap around the new BZ boundary and the filling of the valence band. (d) A visualization of a Mott insulator as represented in the Hubbard model at half-filling. The spheres represent electrons. When the cost $U$ of doubly occupying a site is significantly greater than the bandwidth $W \sim t$ due to electron hopping $t$, a Mott insulating state sets in. (e) DOS plot for a Mott insulator showing the upper and lower Hubbard bands (UHB and LHB).}
\end{figure}

\subsubsection{Mott insulators}
If a material with an odd number of electrons per unit-cell is insulating, the Mott-Hubbard mechanism may be invoked as an explanation. While the Bloch band theory pretends that electrons do not mutually interact, the Hubbard model and its extensions explicitly model the role of electrostatic and magnetic interactions between electrons. Figure 1(d) gives a visualization of the most basic mechanisms in the Hubbard model. The kinetic energy term of the Hubbard Hamiltonian allows for electrons to enjoy energy savings by hopping between lattice sites and forming bands, \textit{via} the orbital overlap parameter $t$. But hopping onto an already occupied site incurs an energy cost $U$ due to electrons' mutual Coulomb repulsion. The band width associated with hopping is $W \sim nt$, where $n$ is the lattice coordination number (e.g. $n = 6$ for a triangular lattice). If, for whatever reason, $W$ is suppressed such that the `Mottness' ratio $U/W$ exceeds some critical value, Coulomb repulsion traps all electrons in a collective insulating state called the Mott state \cite{Mott1937}. This effect opens a Mott gap in the band structure of a system that would otherwise have been metallic due to a partially-filled band crossing the Fermi energy. The remnants of the formerly metallic band that persist above and below the Mott gap are called the upper and lower Hubbard bands [UHB and LHB, see Fig. 1(e)]. The Mott insulating state is of particular interest because it is the `parent' state, from which enigmatic quantum many-body behaviours such as the `strange metal' phase, multiple electronic liquid crystal phases, and high-temperature superconductivity can emerge upon suitable doping \cite{Lee2006,Proust2019}.

\vspace{1pc}
Aside from the kinds of insulator described above, there are a variety of others, including insulators that result from other, more exotic Fermi surface instabilities such as the excitonic insulator, or from other mechanisms such as the charge-transfer insulator (closely related to the Mott insulator), the Anderson insulator \cite{Anderson1958}, the Kondo insulator \cite{Kondo_insulator}, and so on. Particular insulating states may be identified \textit{via} observation of their characteristic excitations, or the differing timescales of each state's dynamics \cite{Hellmann2012}. For example, an insulating state resulting from a purely electronic mechanism, such as a Mott state, exhibits much more rapid dynamics under abrupt perturbation than one that results from some structural or \textit{e-ph} derived mechanism involving the much heavier ions. However, the most fundamental step in identifying an insulating state is to correctly determine the unit-cell of the system (or surface or embedded sub-system) under investigation, and this task will be the subject of a large part of the following discussion.

\subsection{Spin liquids in magnetically frustrated Mott insulators}

As well as an insulating state that localizes electrons on a lattice, another requirement for a spin liquid is that electrons' spins should interact antiferromagnetically. In insulators the interaction between neighbouring spins is mediated by direct exchange, which can result in either ferromagnetism or antiferromagnetism, or by superexchange which tends to favour antiferromagnetism, with an energy saving of $J \sim -t^{2}/U$. Many of the first identified and most intensely investigated Mott insulators are metal oxides with O superexchange centres, and are antiferromagnets \cite{Mott1949,Anderson1963,Lee2006}. Many of these materials, at least in the un-doped `parent' phase, host two-dimensional square lattices with an antiferromagnetic structure depicted in Fig. 2(a). For the putative Mott insulating state of 1$T$-TaS$_{2}$ it is much less clear how neighbouring spins should be expected to interact, and the argument for antiferromagnetic interactions is largely driven by the observed absence of ferromagnetism.

\begin{figure}
\centering
\includegraphics[scale=1]{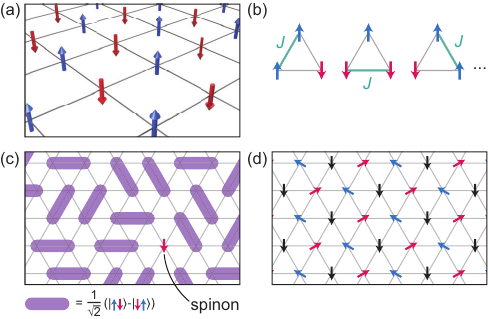}
\caption{\label{fig:2}
Spin liquids in frustrated magnets. (a) Antiferromagnetic order on the square lattice. Each spin's preference to align antiferromagnetically with its neighbours is satisfied in a unique ground state configuration. (b) For lattices with triangular motifs, the spins' preference to anti-align cannot be simultaneously satisfied. An energy cost $\left| J \right|$ is incurred and this can only be re-located, but not disposed of, by flipping any of the spins. This leads to a ground state degeneracy and is the behaviour underpinning classical spin liquids. (c) One possible configuration of spin singlets on a triangular lattice. A QSL is a superposition of all configurations that cover the lattice. A spin that has failed to pair up in a singlet, called a `spinon' excitation of a QSL, is also shown. (d) The likely ground state for antiferromagnetically coupled Heisenberg spins on a triangular lattice, given only nearest-neighbour coupling \cite{Huse1988,Capriotti1989,Sachdev1992,Bernu1992,Zheng2006,White2007}. If next-nearest-neighbour coupling is included, a QSL phase can be recovered \cite{Hu2015,Zhu2015,Iqbal2016,Hu2019, Drescher2023, Sherman2023}.}
\end{figure}

A third requirement for a spin liquid is geometric frustration, which is fulfilled in lattices with triangular motifs, including hexagonal, kagom\'{e} and hyperkagom\'{e} lattices. (The honeycomb lattice does not feature triangular motifs, but can also exhibit long-range frustration, as in the Kitaev model \cite{Kitaev2006}, and even the square lattice can exhibit frustration if beyond-nearest-neighbour interactions are included \cite{Moessner2001}.) Geometric frustration imposes that antiferromagnetic interactions between nearest-neighbours cannot be simultaneously satisfied beyond any two neighbouring spins. As illustrated in Fig. 2(b) for three coupled Ising spins, for example, an energy cost $\left| J \right|$ for parallel alignment must be incurred, and cannot be removed by flipping spins to change between any of six energy-degenerate configurations. For an extended frustrated lattice the same problem leads to a very large number of degenerate configurations and a significant associated entropy \cite{Balents2010}. In the case of a classical spin liquid the microscopic degrees of freedom are the individual spin orientations, and at arbitrarily low temperature the system can fluctuate by flipping spins so as to gradually traverse the large configuration space. In the quantum case, instead, adjacent spins may become entangled by forming singlets in which case the microscopic degrees of freedom are the ways the singlets cover the lattice. This can be captured and analyzed in so-called quantum dimer models \cite{Hwang2015}. One possible configuration is shown for the triangular lattice in Fig. 2(c). In a QSL, due to quantum fluctuations, the system adopts a superposition state of very many such disordered and energy degenerate singlet configurations, which is called a resonating valence bond (RVB) state \cite{Anderson1973}. In this superposition state any local symmetry breaking is smeared out and the QSL adopts the lattice symmetry. A possible excitation of this state is a `spinon', often depicted as a leftover spin that does not belong to a singlet, as shown in Fig. 2(c). More correctly, this is a product of spin-charge separation of electrons into `partons', in this case a spin-carrying but charge-neutral excitation \cite{Anderson1987,Baskaran1987}. It was discovered early on that a RVB state is not in fact the ground state of Anderson's early model for frustrated antiferromagnetism on the triangular lattice (see below). A multitude of different models beyond the early RVB paradigm do host stable QSL ground states \cite{Savary2017}, and many of them are characterized by either a gapped or gapless spectrum of itinerant spinons.

As described above, the QSL is not characterized by any symmetry-breaking or associated local order parameter, which poses a challenge for detection of a QSL state using any local probe. The most direct evidence in favour of a QSL would be provided by non-local properties, such as an entropy contribution characteristic of extended quantum entanglement \cite{Zhou2017,Anderson1973,Jiang2012}. Without a local order parameter as evidence, arguments for the presence of a QSL tend to rely on the absence of magnetic order. Probably the simplest way to observe or exclude magnetic order in a candidate material is by measurement of the magnetic susceptibility which can show the anomaly in susceptibility as order sets in, and establish the Curie-Weiss temperature $\Theta_{\mathrm{CW}}$ that characterizes the strength of magnetic interactions and is negative for antiferromagnets. Careful heat capacity measurements, neutron diffraction and $\mu$-SR observations can also provide evidence for the absence of magnetic ordering. A measure of how resistant a frustrated magnet is to ordering is the frustration parameter $f = \left| \Theta_{\mathrm{CW}} \right| / T_{\mathrm{c}}$ where $T_{\mathrm{c}}$ is the critical temperature at which freezing into a conventional ordered state, if any, is observed. While we have an ordered magnet (or `spin solid') below $T_{\mathrm{c}}$ and a paramagnet (`spin gas') above $ \left| \Theta_{\mathrm{CW}} \right| $, at intermediate temperatures a spin liquid might be realized. Materials with $f \sim 10$ or more are considered to be good candidates for spin liquids \cite{Balents2010}, and 1$T$-TaS$_{2}$ has been estimated to have $f \gtrsim 30$ due to the very low upper bound placed on $T_{\mathrm{c}}$ using muon spin rotation ($\mu$-SR) measurements \cite{Ribak2017}.

Whether a triangular lattice of Heisenberg spins hosts a QSL, even in theory, is unclear. (As an aside, the Kitaev model for the honeycomb lattice is exactly solvable and it is well established that it does host a QSL. For this reason it is currently the most intensely investigated QSL model, but is not the model thought to be realized in triangular lattice compounds such as 1$T$-TaS$_{2}$ and therefore will not be discussed in any detail here.) Analytical and numerical investigations indicate that at $T = 0$ a triangular Heisenberg antiferromagnet does not realize a RVB state, but instead exhibits an ordered ground state comprising three intertwined sub-lattices of parallel spins related by 120$^{\circ}$ rotations \cite{Huse1988,Capriotti1989,Sachdev1992,Bernu1992,Zheng2006,White2007}, as shown in Fig. 2(d). This is true if only nearest-neighbour interactions are considered, but if next-nearest-neighbour interactions are included a regime exists where a QSL state is recovered \cite{Hu2015,Zhu2015,Iqbal2016}, though it is thought to be Dirac spin liquid and not a RVB state \cite{Hu2019, Drescher2023, Sherman2023}.

A further possibility other than a QSL state is a `random singlet' state which can form if, due to quenched disorder, the strength of the exchange interactions $J$ has a distribution with some width. The result is a coverage of the lattice by singlets with a distribution of ranges whose strength greatly diminishes with range, and an overall glass-like configuration of singlets \cite{Kimchi2018a, Kimchi2018b}. This picture was proposed in order to understand the unusual power-law dependence of the magnetic specific heat $C$($T$) in the QSL candidate material YbMgGaO$_{4}$ \cite{Kimchi2018a}, as well as to understand apparent universal $C$[$H$,$T$] scaling seen in a variety of candidate materials that are otherwise seemingly unrelated \cite{Li2019, Li2020}, one of which is 1$T$-TaS$_{2}$ \cite{Murayama2020}. Subsequent results for 1$T$-TaS$_{2}$ have also been suggested to be consistent with this scenario, including the observation of slow relaxation of magnetization said to be consistent with a glassy state slowly switching between many quasi-equilibrium configurations \cite{Pal2020}.

As well as an absence of long range magnetic order, indirect evidence for a QSL can appear in macroscopic thermodynamic measurements which reflect its quasiparticle excitations, and various theoretical models for QSLs each allow their own characteristic excitations. For quantum dimer models on lattices with triangular motifs, the quasiparticle excitations are spinons, and spin-less (i.e. bosonic) charged excitations called `chargons', the latter of which are immobile in a Mott insulator and as such are undetectable. Spinons, being itinerant even in an insulator, can carry heat and contribute entropy. An outcome of this is a linear contribution to the heat capacity as well as deviations from the expected thermal conductivity which in insulators should otherwise come only from phonons. The existence of spinons is well-established in one-dimensional spin chains but is less certain for higher-dimensional systems. In two-dimensional organic salts such as $\kappa$-(ET$_{2}$)Cu$_{2}$(CN)$_{3}$ and EtMe$_{3}$Sb[Pd(dmit)$_{2}$]$_{2}$, evidence has been claimed for spinons despite the absence of magnetic order \cite{Yamashita2008,Yamashita2010}, although this has been challenged by more recent observations \cite{Ni2019, BourgeoisHope2019, Yamashita2020}. Signatures of a spinon contribution to thermal conductivity and heat capacity have also been searched for in 1$T$-TaS$_{2}$ and will be described in the following section.

A merit of some QSL models that feature itinerant spinons is that they support concrete predictions for phenomena that appear in the local electronic DOS and thus are amenable to experimental observation using tunnelling spectroscopy \cite{Tang2013,He2022,He2023a,He2023b}. 1$T$-Ta$X_{2}$ and similar materials are platforms where these predictions can be tested. In Section 6 below we will review the results of recent efforts to do so.

To this date there remains no unambiguous identification of a QSL in any triangular lattice material. Aside from 1$T$-TaS$_{2}$, a number of candidates such as Ba$_{8}$CoNb$_{6}$O$_{24}$, YbZn$_{2}$GaO$_{5}$ and others, that may realize a triangular Heisenberg antiferromagnet with either QSL-like behaviour or interesting related behaviour, have been considered up to very recently \cite{Shen2016,Rawl2017,Cui2018,Rawl2019,Bag2024}.

\section{1$T$ tantalum dichalcogenides}

\begin{figure}
\centering
\includegraphics[scale=1]{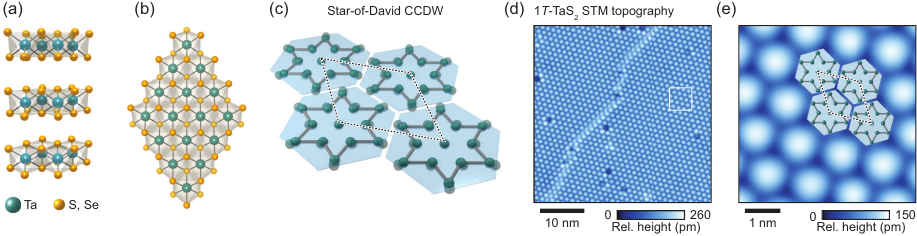}
\caption{\label{fig:3}
(a) A depiction of the quasi-two-dimensional structure of 1$T$-Ta$X_{2}$, and (b) a top-down view, both created using VESTA \cite{VESTA}. (c) An illustration of the formation of the star-of-David (SD) CCDW superstructure showing the displacement of the Ta ions from their positions in the high-temperature phase. The outline of the CDW supercell in the basal plane is shown with a dotted line. (d) A typical STM topography image of the CCDW phase of 1$T$-TaS$_{2}$ ($V$ = -500~mV, $I$ = 250~pA, $T$ = 1.5 K). A domain wall can be seen, characterized by a sharp discontinuity in the phase of the CCDW order parameter. (e) A zoom-in view of the topography in the white square in (d), shown with a depiction of the SD superstructure overlaid.}
\end{figure}

\subsection{The star-of-David charge density wave}

At high temperature, 1$T$-TaS$_{2}$ and 1$T$-TaSe$_{2}$ both have a layered structure of planar triangular lattices of Ta ions each sandwiched between triangular lattices of chalcogen ions, as shown in Figs. 3(a) and 3(b). 1$T$-TaS$_{2}$ has a more complicated behaviour than 1$T$-TaSe$_{2}$ upon cooling, undergoing transitions first to an incommensurate (I-) CDW at $T \approx$ 550~K, a nearly-commensurate (NC-) CDW at $T \approx$ 350~K and finally the commensurate (C-) CDW phase at $T \approx$ 180~K, which for both materials will be the focus of this article. 1$T$-TaSe$_{2}$ enters a CCDW phase upon cooling below $T \approx$ 430~K. For both materials the CCDW phase exhibits an almost completely in-plane periodic distortion of each layer's Ta lattice, described by a $(\sqrt{13} \times \sqrt{13}) R13.9^{\circ}$ supercell. Groups of 13 Ta ions undergo a distortion in which 12 ions contract symmetrically around one located at one of the super-lattice sites, as shown in Fig. 3(c). This forms a lattice of so-called star-of-David (SD) clusters \cite{Wilson1975,Naito1984}. The in-plane distortion of Ta ions is accompanied by a `breathing' distortion of the surrounding chalcogen ions (26 in each supercell), that bow outward at the centre of the cluster. As well as this CDW description, the aforementioned ordered phases can alternatively be described as the successive phases formed from the condensation and crystallization of a fluid of polarons (bound pairs of electrons and local lattice distortions, in this case those forming the SD cluster) that is already present at some low density in the high-temperature phase \cite{Bozin2023}. Some observations made using STM are well situated in this polaron-centric picture \cite{Gerasimenko2019,Ravnik2021}, or even in a quasi-molecular picture of the SD clusters, though below we will maintain the CDW description. Typical images of the SD lattice as seen using low-temperature STM are shown in Figs. 3(d) and 3(e).

The undistorted high-temperature phase has a metallic band structure with a large hole pocket surrounding each M point of the BZ \cite{Ang2012,Yu2017a}. With the onset of the CCDW phase, the BZ shrinks and calculations of the band structure using density-functional theory (DFT) show that the bands undergo a folding that creates a gap in the in-plane propagating bands, interpreted as a CDW gap \cite{Ge2010,Ritschel2015,Martino2020}. There remains one metallic band whose dispersion depends on the arrangement of CDW layers in the out-of-plane direction \cite{Ritschel2015}. Its existence can be understood as a necessity of Bloch band theory. With a supercell that contains one SD cluster and therefore an odd number (13) of Ta valence orbitals, there must be a half-filled band. In detail, the 5$d_{z^{2}}$ orbitals of the 12 Ta ions around the periphery of the SD cluster form covalently bonded pairs, while that of the central Ta ion is left un-paired \cite{Qiao2017}. These are separated from those in adjacent SD clusters of the same layer by the CCDW lattice vector $a_{\mathrm{CDW}} = b_{\mathrm{CDW}} \approx$ 12.1~\AA, which is larger than the inter-layer spacing $c_{0} \approx$ 6~\AA. If adjacent CCDW layers are arranged in an in-phase manner along the out-of-plane direction, it is reasonable to imagine that the largest orbital overlap that can be found is that between Ta $d_{z^{2}}$ orbitals in adjacent layers, and that this results in the formation of an out-of-plane band. If the CCDW layers are arranged in some other way, out-of-phase, the degree of orbital overlap is less clear, but the basic rationale for expecting metallicity due to half-filling remains the same (albeit with different details of the band dispersion \cite{Ge2010,Ritschel2015}).

\begin{figure}
\centering
\includegraphics[scale=1]{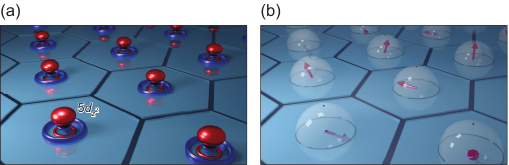}
\caption{\label{fig:4}
(a) Simplified depiction of an array of 5$d_{z^{2}}$ orbitals of central Ta ions in SD clusters in the CCDW phase. These are the orbitals that form the spectral features just above and below the gap in the DOS (the putative UHB and LHB), and typically give the overwhelming contribution to the tunnelling current and resulting STM images [for example Figs. 3(d) and (e)]. (b) The corresponding array of $S = \frac{1}{2}$ spins, depicted here as Heisenberg spins and assuming paramagnetic behaviour at some temperature above that of any QSL or ordered phase.}
\end{figure}

In spite of the above, it is observed that bulk 1$T$-TaS$_{2}$ shows a transition from a metal to an insulator upon formation of the CCDW \cite{Fazekas1980}. (Bulk 1$T$-TaSe$_{2}$ on the other hand does not, an important fact that we will return to below.) From this it has often been inferred that the Mott-Hubbard mechanism must be invoked as an explanation \cite{Fazekas1980,Fazekas1979}. Affirmative evidence for this has been put forward in recent years, in the form of ultra-fast dynamics seen in time-resolved optical reflectivity and photoemission measurements \cite{Hellmann2012,Perfetti2008,Petersen2011,Stojchevska2014,Mann2016,Ligges2018,Zhang2019,Avigo2020}. Accepting the Mott-Hubbard picture and disregarding inter-layer effects for now, the system can be thought of as an array of un-paired Ta $d_{z^{2}}$ orbitals localized on the triangular lattice of SD centres, as illustrated in Fig. 4(a). These each carry spin $S = \frac{1}{2}$ spins, as illustrated in Fig. 4(b). This scenario is supported by neutron scattering and other measurements that were used to estimate a magnetic moment of $\sim$0.4~$\mu_{\textrm{B}}$ per SD cluster \cite{Kratochvilova2017}.

\begin{figure}
\centering
\includegraphics[scale=1]{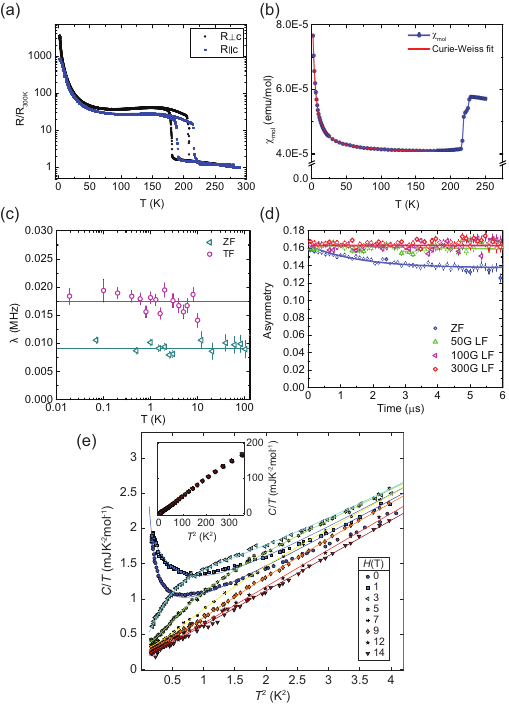}
\caption{\label{fig:5}
(a) Resistance and (b) susceptibility data for bulk 1$T$-TaS$_{2}$ reported by Ribak \textit{et al.} \cite{Ribak2017}. (c) Results of $\mu$-SR measurements under zero field (ZF) and a weak transverse field (TF), showing that upon cooling to $\sim$20~mK, there is no significant increase of the muon polarization damping rate $\lambda$ which would indicate freezing into a magnetically ordered state. (d) Flat $\mu$-SR asymmetry under various longitudinal fields (LF), consistent with the presence of only nuclear moments. (e) Heat capacity measurements showing deviations at low temperature from the expected phonon contribution $C_{p} \propto T^{3}$. These deviations might result from spinon excitations of a QSL. Reproduced with permission from \cite{Ribak2017}. Copyright of the American Physical Society.}
\end{figure}

\subsection{Macroscopic evidence for and against a QSL formed from SD cluster spins}

The triangular lattice of localized spins was recognized by Law and Lee \cite{Law2017} to be a potential host for a spin liquid phase. Since early on, no long range magnetic order was observed in 1$T$-TaS$_{2}$ \cite{Fazekas1980}, and more recently $\mu$-SR measurements have shown an absence of any conventional magnetic order down to temperatures as low as a few tens of milli-Kelvins \cite{Kratochvilova2017,Ribak2017,Klanjsek2017} (see Fig. 5). Magnetic susceptibility measurements \cite{Ribak2017} have shown a large temperature independent contribution to the magnetic susceptibility consistent with paramagnetism, and Curie-Weiss behaviour with a Curie constant that would account for no more than $\sim$10~\% of the expected number of unpaired spins in the sample. A possible interpretation was suggested to be the delocalization of upwards of $\sim$90~\% of the spins, though another possible explanation is the predominance of localized singlet formation as in a valence bond solid.

Contributions to the heat capacity and specific heat from spinons have also been searched for. One report on the thermal conductivity $\kappa$ showed no itinerant fermionic magnetic excitations (such as spinons) in 1$T$-TaS$_{2}$ down to 0.1~K \cite{Yu2017b}. In contrast, a later report indicated a linear contribution to $\kappa / T$ with a finite residue as $T$ approaches zero, consistent with the presence of gapless spinons \cite{Murayama2020}. Nuclear quadrupole resonance indicated gapless spinons only between $T$ = 200~K and 55~K, but an amorphous tiling of frozen singlets emerging from the putative QSL below 55~K \cite{Klanjsek2017}. Other $\mu$-SR and polarized neutron scattering measurements indicated short ranged magnetic order emerging below $T$ = 50~K \cite{Kratochvilova2017}.

Although the above results tend to indicate antiferromagnetic coupling between the spins of the SD clusters (negative Curie-Weiss temperature), but without long range antiferromagnetic order, an interesting recent theoretical work suggested that the coupling within a single 1$T$-TaS$_{2}$ layer might be ferromagnetic \cite{Pasquier2022}, with significant magneto-crystalline anisotropy that leads to in-plane magnetization. A possible reason that overall ferromagnetic order is then not observed was suggested to be inter-layer coupling. Indeed, here it is worth noting that most of the work done in search of QSL behaviour in bulk 1$T$-TaS$_{2}$ was done when less was known about inter-layer interactions. Not only the fate of the SD clusters' spins in the presence of inter-layer coupling, but also the status of bulk 1$T$-TaS$_{2}$ as a Mott insulator may be profoundly impacted by inter-layer effects, as we discuss below.


\subsection{Inter-layer stacking}

An important step in understanding any given electronic band structure is the reconciliation of observed band dispersions, as obtained using angle-resolved photo-emission spectroscopy (ARPES) for example, and the band structure predicted using DFT calculations and comparable methods. A DFT treatment of the bulk or near-surface band structure in the CCDW state necessarily makes some assumption about the way layers of SD clusters are arranged along out-of-plane axis, i.e. the stacking of the CCDW layers. The stacking of two adjacent layers can be described with a vector \textbf{t} which, due to the condition of commensurability, must have in-plane components that are a linear combination of the in-plane lattice vectors $\mathbf{a}_{0}$ and $\mathbf{b}_{0}$ of the un-distorted structure, i.e. $\mathbf{t} = m \mathbf{a}_{0} + n \mathbf{b}_{0} + \mathbf{c}_{0}$.

Notation used to identify various layer-layer stacking configurations in the CCDW of 1$T$-TaS$_{2}$ differs, but we try to reconcile the popular conventions in a single framework here. When overlaying one SD layer on another, the central Ta ion of the upper SD cluster can align with any of 13 Ta sites in the layer below. The sites can be labeled with an index $i \in \{ 0 \ldots 12 \}$ (or alternatively $i \in \{ A \ldots L \}$ in an often-used convention \cite{Lee2019}), with 0 (or $A$) referring to the central site, as shown in Fig. 6(a). The stacking vector, which connects the central Ta ions of SD clusters in adjacent layers, is then $\mathbf{t}_{i}$ \cite{Nakanishi1984,Ishiguro1991,Ritschel2018}. Given the three-fold symmetry of the SD pattern [recalling the octahedral coordination of S ions around each Ta, neglected in Fig. 5(a) for clarity], the 13 stacking vectors can be grouped into five sets: $\mathbf{t}_{0}$ for on-top stacking, $\mathbf{t}_{ \{ 1,3,9 \} }$ and $\mathbf{t}_{ \{ 4,10,12 \} }$ [black and gray ions in Fig. 6(a)], and $\mathbf{t}_{ \{ 2,5,6 \} }$ and $\mathbf{t}_{ \{ 7,8,11 \} }$ [red and pink ions in Fig. 6(a)] \cite{Ritschel2018}. The CCDW configuration has two possible instantiations related by a mirror symmetry operation, for example through a $\langle 010 \rangle$ plane of the un-distorted structure \cite{Zong2018,Sung2022}. One or the other of these may be observed in any given microscopic measurement \cite{Qiao2017,Ma2016}. Therefore the sets $i \in \{ 1,3,9 \}$ and $i \in \{ 4,10,12 \}$ as defined in Fig. 6(a) are interchangeable, as are the sets $i \in \{ 2,5,6 \}$ and $i \in \{ 7,8,11 \}$. (They are uniquely defined only when the S lattice is included in the diagram.)

In STM measurements the distinction between, for example, the $i \in { \{ 2,5,6 \} }$ and $i \in { \{ 7,8,11 \} }$ sites is usually ambiguous. The S lattice must be resolved in order to uniquely identify them, and this resolution is only rarely possible, seeming only to be available if the STM tip apex suffers from some contamination. (In such cases it can even be unclear whether the atom locations correspond to the local maxima or minima of topographic image.) Given these complications, it is useful to define three sets encompassing the stacking vectors as follows: $\mathbf{T}_{A} = \{ \mathbf{t}_{0} \}$, $\mathbf{T}_{B} = \{ \mathbf{t}_{ \{ 1,3,9 \} }, \mathbf{t}_{ \{ 4,10,12 \} } \}$, and $\mathbf{T}_{C} = \{ \mathbf{t}_{ \{ 2,5,6 \} }, \mathbf{t}_{ \{ 7,8,11 \} } \}$. The previous, more detailed notation is necessary to fully describe extended out-of-plane chains of stacking \cite{Ishiguro1991}, but because STM measurement have so far only been able to observe at most two or three layers of SD clusters, they have only been able to characterize at most two consecutive stacking vectors, and the $\mathbf{T}_{A,B,C}$ notation suffices to describe all STM observations so far.

\begin{figure}
\centering
\includegraphics[scale=1]{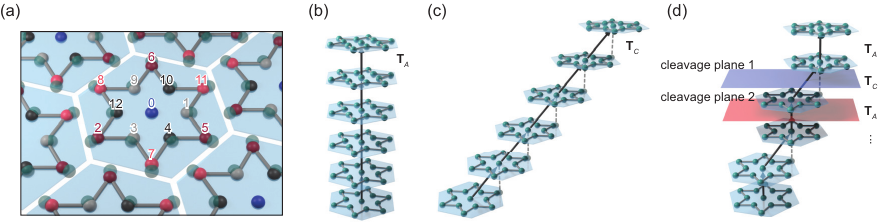}
\caption{\label{fig:6}
(a) Ta sites in a SD cluster as labelled in the convention used by Ritschel \textit{et al.} \cite{Ritschel2018}. (b) Depiction of an out-of-plane chain of SD clusters with a uniform $\mathbf{T}_{A}$ stacking pattern. (c) An out-of-plane chain with a uniform $\mathbf{T}_{C}$ stacking pattern (as is realized in bulk 1$T$-TaSe$_{2}$). (d) The alternating $\mathbf{T}_{A}$, $\mathbf{T}_{C}$, $\mathbf{T}_{A}$... stacking pattern (henceforth `\textit{AC}') proposed by Naito \textit{et al.} \cite{Naito1986}. This structure features two in-equivalent cleavage planes, labelled here as planes 1 \& 2, which may have very different electronic properties if inter-layer interactions are significant.}
\end{figure}

When constructing a supercell in order to predict the electronic structure, the simplest assumption is that of `on-top' stacking, $\mathbf{T}_{A}$, shown in Fig. 6(b). Using DFT (without electron-electron interaction) this was shown to yield a large gap for all in-plane momenta, but leaves a single band that crosses the Fermi level between $\Gamma$ and A. In this case the system can be thought of as an in-plane CDW insulator and a quasi-one-dimensional out-of-plane metal \cite{Ritschel2015,Darancet2014,Ngankeu2017,Yu2017a}. It is then only the observed lack of out-of-plane metallicity that demands to be understood. For uniform $\mathbf{T}_{C}$ stacking, shown in Fig. 6(c), bands cross the Fermi level at multiple purely in-plane as well as out-of-plane momenta. Even if an effort is made to include an on-site Coulomb repulsion $U$ to mimic some effects beyond non-interacting band theory, neither simple $\mathbf{T}_{A}$ nor $\mathbf{T}_{C}$ stacking leads to satisfactory match with the valence band dispersion observed using ARPES \cite{Ritschel2015}.

It had been recognized early on, \textit{via} x-ray and nuclear quadrupole resonance investigations, that the stacking is not simple or regular but instead probably exhibits a disordered sequence of stacking vectors \cite{Naito1984,Tanda1984}. Here the distinction between the sets of stacking vectors $\mathbf{t}_{ \{ 2,5,6 \} }$ and $\mathbf{t}_{ \{ 7,8,11 \} }$ is important, because it seems that where the stacking features disorder, stacking vectors are randomly drawn from either $\mathbf{t}_{ \{ 2,5,6 \} }$ or $\mathbf{t}_{ \{ 7,8,11 \} }$ \cite{Ishiguro1991}, which itself suggests that the free energy is sensitive to beyond-nearest-neighbour stacking environment of any given layer.

In a prescient paper, Naito \textit{et al.} \cite{Naito1986} suggested un-paired 5$d$ orbitals in SD clusters could form covalent bonds crossing the van der Waals gap between layers, such that adjacent layers prefer in-phase (on-top) stacking, and the overall structure would be composed of some stacking pattern of these on-top-stacked bilayer elements. Naito \textit{et al.} further suggested that as well as an attractive \textit{intra}-bilayer coupling, there should be a repulsive \textit{inter}-bilayer coupling, so that while the intra-bilayer stacking would be in-phase, and the inter-bilayer stacking would be as close to anti-phase as possible. Thus the alternating pattern of $\mathbf{T}_{A}$ and $\mathbf{T}_{C}$ stacking was first proposed, which we will refer to below as the $AC$ stacking pattern. This is illustrated in Fig. 6(d). This was later supported by early transmission electron microscopy observations suggesting a disordered stacking pattern in which nearly every alternate stacking vector was chosen from $\mathbf{t}_{ \{ 7,8,11 \} } \in \mathbf{T}_{C}$ \cite{Ishiguro1991}.

More recently Ritschel \textit{et al.} investigated the consequences of such an alternating stacking structure using DFT, demonstrating that it predicts a filled valence band and complete insulating gap for stacking that alternates in a long-range-ordered manner between $\mathbf{t}_{0}$ and $\mathbf{t}_{2}$, as shown in Fig. 7 \cite{Ritschel2018}. Importantly, this prediction of insulating behaviour was made without any consideration of electronic correlations. This conclusion was echoed by Lee \textit{et al.} \cite{Lee2019} who considered stacking that alternates between $\mathbf{t}_{0}$ and $\mathbf{t}_{7}$ (in Ritschel's notation).

\begin{figure}
\centering
\includegraphics[scale=1]{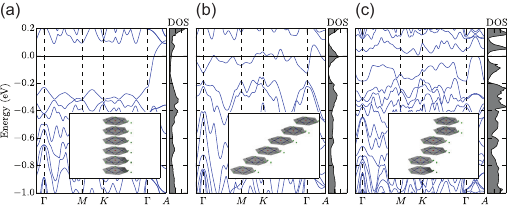}
\caption{\label{fig:7}
Band structures calculated by Ritschel \textit{et al.}, using DFT, for (a) uniform $\mathbf{T}_{A}$ stacking, (b) uniform $\mathbf{T}_{C}$ stacking and (c) the alternating $AC$ stacking pattern. Both of the uniform stacking patterns yield a metallic band, while the alternating stacking, with two SD cluster per cell, yields a filled valence band and a gap throughout the BZ. Reproduced with permission from \cite{Ritschel2018}. Copyright of the American Physical Society.}
\end{figure}

Experimental results using x-ray diffraction \cite{Stahl2020}, high- and low-energy electron diffraction techniques \cite{LeGuyader2017,vonWitte2019}, angle-resolved photo-emission spectroscopy \cite{Wang2020}, nuclear magnetic resonance observations \cite{Cheng2023}, and combined STM and tunnelling spectroscopy \cite{Butler2020}, have bolstered the view that an alternating stacking is indeed realized in the CCDW phase of 1$T$-TaS$_{2}$, though there have also been observations of multi-domain structures realizing a variety of other stacking patterns \cite{Wang2023}. What is the driving mechanism of the layer-wise pairing? Because simple band theory would predict that uniform $\mathbf{T_{\mathrm{A}}}$ stacking would yield an out-of-plane one-dimensional metal, it is tempting to invoke the Peierls instability and posit that the uniform $\mathbf{T_{\mathrm{A}}}$-stacked metal would destabilize towards a layer-wise dimerized insulator. However, a Peierls instability is characterized by a continuous order parameter, which is not possible in the three-dimensional CCDW structure due to the discrete nature of the stacking vectors.

Regardless of the driving mechanism behind the formation of the bilayer correlation, its consequence is that the supercell of the three-dimensional CCDW order encompasses two SD clusters and therefore an even number of orbitals. Therefore a band theoretic treatment of the electronic structure should be expected to find a filled valence band and there is no need to resort to the Mott-Hubbard mechanism in order to explain the insulating behaviour. The bilayer correlation also potentially precludes the possibility of a QSL. If the central Ta $d_{z^{2}}$ orbitals of each SD cluster form stable dimers between layers, the system no longer forms a QSL. Referring to the na\"{i}ve picture of Fig. 2(c), instead of the RVB state, the system would adopt a single unique ground state called a valence bond solid.

Although a bilayer correlation seems to indicate a band insulator and to disfavour a QSL, strong electronic correlations may still be present and may be very significant in determining the electronic properties. The aforementioned evidence for a correlated insulating state from ultra-fast dynamics cannot be dismissed, and in Section 4 below, we will see that a detailed assessment of STM results gives a compelling argument that electronic-correlations are indispensable for a comprehensive understanding of the insulating state, even in the presence of bulk dimerization.

First we briefly return to the bulk CCDW of 1$T$-TaSe$_{2}$. It is fairly well established that it exhibits a uniform $\mathbf{T}_{C}$ stacking \cite{Wiegers2001}, depicted in Fig. 6(d), which would yield an odd number of electrons per cell and would naturally explain why it is found to show metallicity in transport measurements \cite{DiSalvo1974}. This is itself noteworthy, since if the Mott-Hubbard argument applies for insulating behaviour of 1$T$-TaS$_{2}$, this invites an explanation for why it does not similarly apply in the case of 1$T$-TaSe$_{2}$. It also reinforces that the comparative richness of inter-layer stacking seen in 1$T$-TaS$_{2}$ is likely the key to why it undergoes a metal-insulator transition upon cooling into the CCDW phase.

\section{STM observations at the surface of bulk 1$T$-TaS$_{2}$}

Low-temperature STM measurements have been able to both observe the CCDW and measure tunnelling spectra at the surface of 1$T$-TaS$_{2}$ for several decades \cite{Kim1994}. A notable resurgence of STM reports from around 2016 focused on microscopic mechanisms leading to local metallicity. These coincided with renewed interest in the insulator-metal transitions of 1$T$-TaS$_{2}$ under external perturbation (electric current pulses \cite{Yoshida2015,Tsen2015}, laser excitation \cite{Stojchevska2014,Vaskivskyi2015,Sun2018}, or pressure \cite{Wang2018} for example). STM reports described the manipulation of the CCDW structure using local electric field pulses to create domain wall networks within which a correlated metal phase was found \cite{Ma2016,Cho2016}. Either the inter-layer stacking degrees of freedom \cite{Ma2016}, or a loss of long-range coherence of the CDW order parameter \cite{Cho2016}, were proposed to be involved in the mechanism of these transitions. Further work examined the potential metallicity and other electronic properties of the domain walls themselves \cite{Cho2017,Park2021}. (These, and contemporaneous works, implicitly adopted the Mott-Hubbard explanation for the insulating nature of the un-perturbed states \cite{Qiao2017}.)

\subsection{Distinct surface terminations of the bilayer-correlated stacking pattern}

Following these developments, our work \cite{Butler2020} addressed the observable consequences in STM measurements given a non-simple inter-layer stacking like the $AC$ stacking pattern. Key to this was the recognition that a three-dimensional structure composed of bilayers necessarily forms either of two distinct types of surface upon cleavage [recall Fig. 6(d)]. A cleave can occur in the gap between two bilayers, leaving an intact bilayer at each of the newly created surfaces, or it can split a bilayer, leaving an un-paired layer at each surface. Tunnelling spectroscopy measurements performed at each one of a relatively large number (24) of cleaved surfaces resulted in DOS spectra that clearly fell into two categories distinguished by the size of the gap in the DOS. These categories are exemplified by the curves shown in Fig. 8. Three quarters of the samples exhibited a gap of $\sim$150~meV, while a quarter had a smaller gap of $\sim$50~meV and an apparent splitting of the features usually identified as the UHB and LHB.

\begin{figure}
\centering
\includegraphics[scale=1]{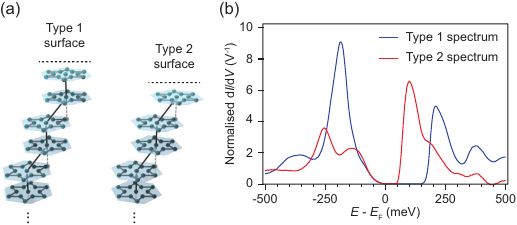}
\caption{\label{fig:8}
(a) The Type 1 and Type 2 surface terminations that can be formed by cleavage at planes 1 and 2 in the $AC$ stacking pattern. (b) Examples taken from two categories of tunnelling conductance spectra that can be measured at surfaces resulting from low-temperature cleavage of 1$T$-TaS$_{2}$. The association between the types of surface and the resulting spectra is proposed in \cite{Butler2020}, as described below. This bifurcated behaviour has been reproduced in a number of subsequent STM reports \cite{Lee2021,Wu2022,Park2023a,Yang2024}. Reproduced from \cite{Butler2020} under CC BY 4.0 license.}
\end{figure}

As the Type 1 surface preserves the dimerized bulk behaviour, it can na\"{i}vely be understood as a filled band insulator without resorting to the Mott-Hubbard mechanism. The expected behaviour for the Type 2 surface is less clear. If we think of the Type 2 surface as an independent two-dimensional system overlaid on top of a bilayer, it must be half-filled and should therefore be expected to be a metal. The likelihood that significant inter-layer coupling complicates this situation cannot be ignored, but nevertheless, DFT calculations that model the Type 1 and 2 surfaces in the absence of electron-electron interactions show that a narrow metallic band should be expected for the Type 2 surface \cite{Jung2022,Li2022}. In fact neither surface is observed to be metallic. Even before identifying which measured spectrum results from which type of surface, we are immediately forced to resurrect the Mott-Hubbard mechanism to explain the fact that both surfaces show insulating behaviour.

Further evidence in favour of the presence of bilayer stacking is obtained from the observation, shown in Fig. 9, of a step-terrace formation where three terraces could be observed. Here the spectrum was seen to alternate between the two types (albeit only for a very short sequence of only three layers), and this is consistent with the cleavage planes throughout a stack of bilayers alternating between planes 1 and 2, leading to alternating Type 1 and 2 surface terminations. In more limited measurements of only two terraces, the type of spectrum was seen to change from one to the other, also consistent with the bilayer pattern \cite{Butler2020_SI}.

\begin{figure}
\centering
\includegraphics[scale=1]{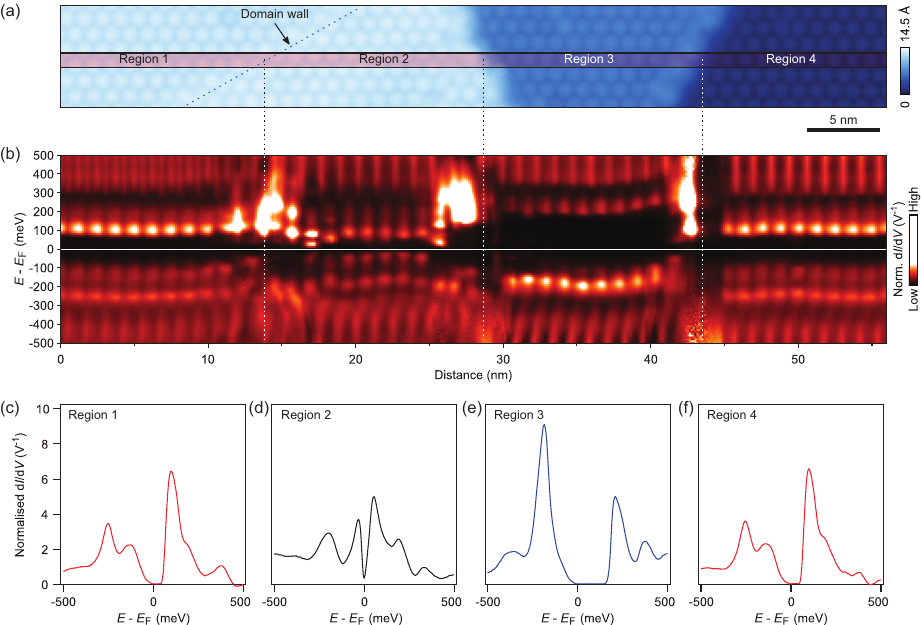}
\caption{\label{fig:9}
(a) STM topography at a step-terrace formation found at the low-temperature-cleaved surface of 1$T$-Ta$_{2}$. Three terraces and one domain wall are seen, resulting in four distinct regions. (b) Tunnelling spectroscopic linecut across the four regions. (c--f) Representative tunnelling spectra from each region. Spectra from the 1$^{\mathrm{st}}$ and 3$^{\mathrm{rd}}$ terraces (Regions 1 and 4) fall into the category of Type 2, while the spectrum from the 2$^{\mathrm{nd}}$ terrace is of Type 1. The spectrum from Region 2 resembles that of a correlated metal and does not fit into either category. Adapted from \cite{Butler2020} under CC BY 4.0 license.}
\end{figure}

\subsection{Identification of layer-wise paired and un-paired surface terminations}

Further analysis of observations on the step-terrace morphology lead to an explicit identification of which location (and which obtained DOS spectrum), corresponds to the Type 1 (or 2) surface. This is achieved by observing the in-plane shift between the SD lattices in adjacent layers, as shown in Fig. 10. If cleavage changes from plane 1 to plane 2 at a step-edge, the upper and lower terraces are implied to belong to the same bilayer in which the CCDW are arranged in-phase and there is no in-plane phase discontinuity in the CDWs. The upper terrace is terminated by the upper member of the dimerized pair and the other member is found un-paired on the lower terrace. On the other hand, if cleavage changes from plane 2 to plane 1, the upper terrace is an un-paired layer lying atop a complete pair, in which case the in-plane shift seen across the step-edge is expected to be the in-plane projection of $\mathbf{T}_{C}$. We can now identify the large (small) gap spectrum with the (un-)paired termination. This suggests the interpretation that the larger gap of the Type 1 surface is a simple band gap, while the smaller gap of the Type 2 surface is a Mott gap. But at a minimum, this observation clearly shows that the stacking of the surface layer atop the underlying layers has a profound effect on its electronic structure, meaning that inter-layer coupling is significant in 1$T$-TaS$_{2}$ and that the microscopic degrees of freedom of inter-layer stacking are therefore indispensable in any understanding of the CCDW ground state.

At this point it is worth noting that in these and more recent measurements \cite{Wu2022}, the paired surface was observed several times more often than the un-paired surface, although these observations are somewhat lacking in statistical power and are only suggestive. A possible interpretation is that cleavage that preserves dimers is significantly energetically favoured over cleavage that breaks them, leading to the Type 1 surface occurring more often. It is also interesting to consider why the observation of two distinct surface terminations of the CCDW structure had not been observed (or at least reported) earlier in the decades-long history of low-temperature STM measurements on 1$T$-TaS$_{2}$ \cite{Kim1994}. (All published DOS spectra prior to 2020 resemble the spectrum we call Type 1.) A likely explanation is related to the order of CCDW formation and sample cleavage. The configuration of the CCDW at surfaces may be different if the sample is first cooled and then cleaved, or first cleaved and then cooled. If the sample is cooled into the CCDW phase and cleaved at low temperature, the bulk-like terminations of the three-dimensional CCDW structure may become accessible to measurement (provided that the sample is not warmed out of the CCDW phase during transfer into the microscope). On the other hand, if the sample is cleaved at room temperature and subsequently cooled upon insertion into a cold microscope, which is much more common, the CCDW might nucleate at surfaces with a preferred configuration (implied to be the dimerized configuration). This provides a possible explanation for why the Type 2 surface went un-observed for so long.

\begin{figure}
\centering
\includegraphics[scale=1]{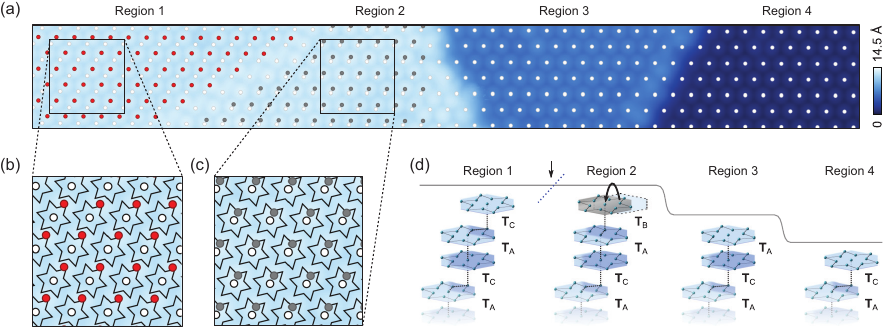}
\caption{\label{fig:10}
(a) CCDW lattice registry across three observed terraces. Regions 3 and 4 are found to be in good registry, and are therefore inferred to belong to the same bilayer. The array of white dots represents the SD lattice of Regions 3 and 4, and this can be extended to Regions 1 and 2 to infer the stacking vector with respect to the underlying bilayer (the continuation of the terrace seen in Region 3). Atop this bilayer, Regions 1 and 2 are found to exhibit $\mathbf{T}_{C}$ and $\mathbf{T}_{B}$ stacking, respectively. This indicates that the uppermost layer in Region 1 has a stacking native to the bulk structure but is now un-paired, and that Region 2 (a correlated metal) has a stacking that is foreign to the bulk stacking order, perhaps due to extrinsic effects during cleavage. Adapted from \cite{Butler2020} under CC BY 4.0 license.}
\end{figure}

As well as the in-plane shifts of the CDW between adjacent terraces that were expected given the proposed $AC$ stacking pattern, an in-plane shift was also observed that was not expected. Instead, in a small region near one of the steps, an in-plane shift was observed that corresponds to $\mathbf{T}_{B}$ stacking of an un-paired layer atop an underlying bilayer. This region was observed to have a metallic DOS spectrum, and this provides an additional clue for understanding the microscopic mechanisms underpinning the metal-insulator transitions of the CCDW phase.

The identification of the paired and un-paired surface terminations of the CCDW was later supported by successful matches between the observed tunnelling spectra and the DOS spectra predicted for the paired and un-paired layers using dynamical mean field theory (DMFT) calculations, which attempt to model the behaviour of strongly correlated electron systems where DFT fails \cite{Georges1996}. The resulting calculated DOS spectra and a compelling comparison with measurement results are shown in Fig. 11 \cite{Petocchi2022}.

\begin{figure}
\centering
\includegraphics[scale=1]{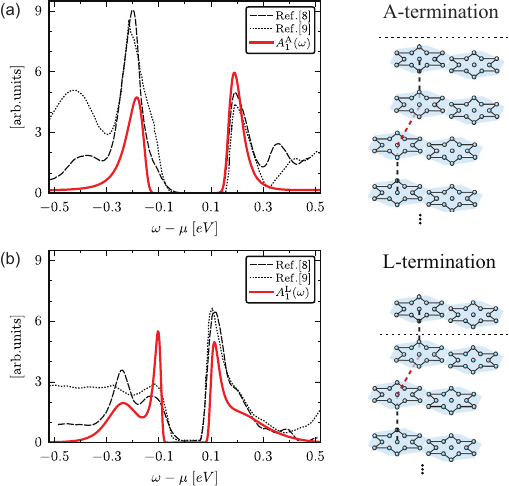}
\caption{\label{fig:11}
A comparison, by Petocchi \textit{et al.}, of DOS spectra calculated using DMFT (red curves), versus corresponding measured tunnelling spectra (black dotted and dashed curves), for (a) the Type 1 surface and (b) the Type 2 surface. Reproduced with permission from \cite{Petocchi2022}. Copyright of the American Physical Society.}
\end{figure}

As described above, while the band theoretic understanding of the insulating gap suffices for the Type 1 surface, it seems necessary to resort to the Mott-Hubbard mechanism to understand the gap at the Type 2 surface. In this case, one can imagine that the CCDW state of 1$T$-TaS$_{2}$ is a bulk band insulator that sometimes (depending on cleavage) has a two-dimensional Mott insulator overlaid on its surface. Experimental evidence supporting this view was put forward in an elegant experiment by Lee \textit{et al.} \cite{Lee2021}, exploiting the different behaviours expected for band and Mott insulators upon doping with electron donors. As shown in Fig. 12, upon sparse deposition of K on cleaved surfaces, the spectrum measured at the Type 1 surface was seen to undergo a rigid shift expected for the doping of an ordinary semiconductor, while the Type 2 surface showed the suppression of the UHB characteristic of a Mott insulator.

\begin{figure}
\centering
\includegraphics[scale=1]{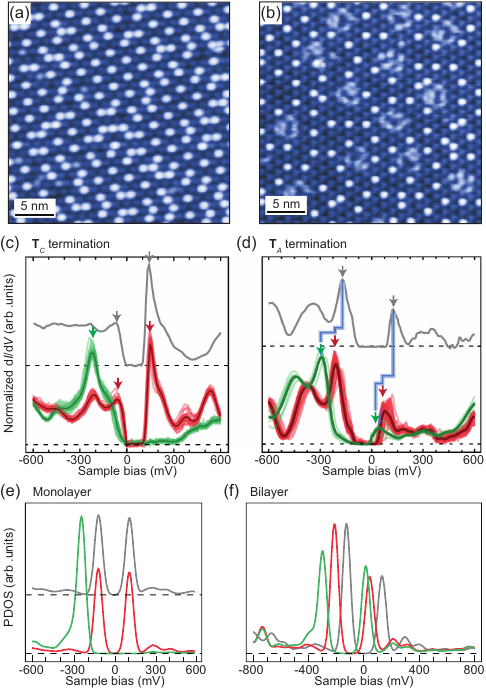}
\caption{\label{fig:12}
Effects of K doping on the two surface terminations of the bilayer correlated CCDW. (a) and (b) STM topography of Type 2 and Type 1 surfaces, respectively, after K deposition. (c) Tunnelling spectra at K-adsorbed sites (green curves) and non-adsorbed sites (red curves), for comparison with the a tunnelling spectrum of the pristine Type 2 surface (gray). (d) Corresponding spectra for the Type 1 surface. (e) and (f) Calculated DOS spectra for a monolayer (i.e. Mott insulator) and bilayer, which match the observed behaviours well. Spectra after doping and before doping are shown in green and red respectively. While the bilayer exhibits a doping effect similar to that of a semiconductor, the monolayer exhibits behaviour expected for a Mott insulator, namely suppression of the UHB and enhancement of the LHB. Reproduced with permission from \cite{Lee2021}. Copyright of the American Physical Society.}
\end{figure}

An additional piece of circumstantial evidence for a greater degree of Mottness at the Type 2 surface can be derived from observations of \textit{e-ph} interactions in STM experiments. In spectroscopy measurements (not limited to STM \cite{Ishizaka2008}) replicas of a peak in the DOS can sometimes be generated by the coupling of the electronic state to an optical phonon. This is described quantitatively in the Franck-Condon scheme \cite{TokmakoffOpenCourseWare}, where the relative intensities among the series of replicas are controlled by the Huang-Rhys parameter $D$ that represents the \textit{e-ph} coupling strength. The greater $D$ is, the greater the number of replicas that can practically be observed. In the context of STM measurements this effect has been described extensively for spectroscopy of individual molecules on surfaces, where the spectrum exhibits replicas of a molecular orbital due to coupling with a molecular vibron \cite{Wu2004,Qiu2004,Nazin2005}. In the case of tunnelling spectroscopy of the 1$T$-TaS$_{2}$ CCDW, upon increasing the bias from 0, the first electron from the tip tunnels into $d_{z^{2}}$ orbital of the central Ta atom of the SD cluster [recall Fig. 4(a)], where the charge density is already at its highest. The arriving electron adds to the charge density and excites the phonon associated with the CDW's periodic lattice distortion. The excitation of this `amplitude mode' phonon has been observed in time-resolved reflectivity and photo-emission measurements \cite{Perfetti2008,Perfetti2006,Kusar2011} that similarly abruptly move electrons into the lowest unoccupied state located at the SD cluster centre. A series of Franck-Condon replicas was seen in STM measurements on the Type 2 surface \cite{Butler2021}, in which a sharp spectroscopic feature (to be discussed again below) was accompanied by replica peaks with a spacing closely matching the energy of the amplitude mode as determined using Raman spectroscopy \cite{Hangyo1983,Albertini2016,Hu2018,Ruggeri2024}. However, for this, the excess electron injected by the STM tip must be localized in a single SD cluster for a sufficiently long time for the \textit{e-ph} coupling to take effect (comparable to one period of the amplitude mode oscillation). It is observed that $D$, the \textit{e-ph} coupling strength, is significantly higher at the Type 2 surface. This can be interpreted as an enhanced doublon lifetime (-- a `doublon' being the quasiparticle associated with a doubly occupied site in the Hubbard model --), which scales exponentially with the Mottness ratio $U/W$ \cite{Sensarma2010}.

\subsection{Stacking-controlled insulator-metal transitions}

Earlier STM works reported that a local electric field pulse applied using the STM tip could drive the 1$T$-TaS$_{2}$ CCDW through a insulator-metal transition \cite{Ma2016,Cho2016}, and Ma \textit{et al.} argued that inter-layer stacking allowed the crucial degrees of freedom for this transition to occur \cite{Ma2016}.
The findings summarized above help to support this argument, and offer a more detailed understanding of the mechanism of the transition. The $\mathbf{T}_{B}$- and $\mathbf{T}_{C}$-stacked surface layers seen in Figs. 9 and 10 above can each be thought of as a two-dimensional system that in absence of electron-electron interactions should have a half-filled band, atop an underlying bilayer. This invites the question of why, despite such a small difference in the stacking configuration, one is observed to be a metal while the other is apparently Mott insulating. The inter-layer hopping between $d_{z^{2}}$ orbitals in adjacent layers, which we may denote as $t_{\perp}$, may be assumed to depend very sensitively on the in-plane displacement between the layers.
Indeed DFT calculations presented by Jung \textit{et al.} show a roughly two-fold increase in the bandwidth of the expected metallic band for the $\mathbf{T}_{B}$-stacked surface as compared to the $\mathbf{T}_{C}$-stacked one \cite{Jung2022}. This calculated bandwidth can be associated loosely with $W$ in the Mottness ratio $U/W$, and this indicates that the insulator-metal transition that occurs upon changing from a $\mathbf{T}_{C}$-stacked surface layer to a $\mathbf{T}_{B}$-stacked surface layer could be a bandwidth-mediated Mott transition.

\subsection{Surface reconstructions of the SD crystal}

Following the association of Type 1 and 2 spectra with cleavage planes 1 and 2, described above, a report by Wu \textit{et al.} drew attention to apparently contradictory observations in similar STM measurements at step-terrace formations \cite{Wu2022}. Although they reproduced all the observations made in prior reports, they also found that a Type 1 spectrum can sometimes appear regardless of the apparent stacking between the uppermost two layers. The interpretation was that the large gap spectrum reflects the intrinsic electronic structure for a CCDW single layer, and that this might occasionally be perturbed towards a small-gap or metallic state due to some unspecified surface or edge phenomenon, or a hidden degree of freedom other than stacking. The proposal that the large gap spectrum reflects the intrinsic single-layer behaviour implies that the gap must be a Mott gap in all cases, and this has some merit when we also consider insulating behaviour seen in isolated 1$T$-TaS$_{2}$ layers in some measurements, which we will discuss below.

\begin{figure}
\centering
\includegraphics[scale=1]{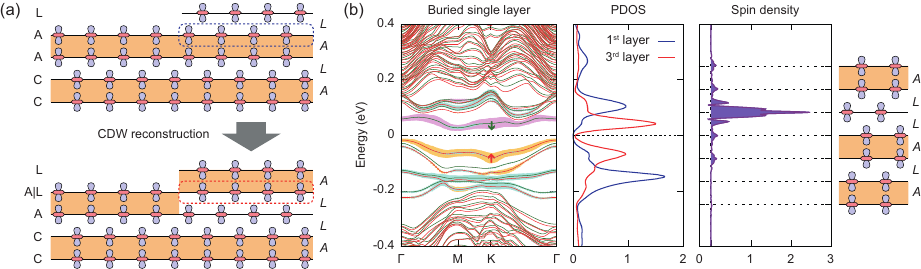}
\caption{\label{fig:13}
(a) Depiction of a surface reconstruction at the Type 2 surface of the bilayer-stacked CCDW that brings layer-wise pairing to the surface, leaves a buried layer un-paired, and creates a domain wall just below the step-edge. (b) Electronic structure, DOS spectrum, and local spin density for a buried un-paired layer amid the bilayer stacking pattern. Reproduced from \cite{Lee2023} under CC BY 4.0 license.}
\end{figure}

The proposed challenge to the $AC$ stacking picture was addressed by Lee \textit{et al.} who suggested, using DFT, that after formation of a bulk like termination of the $AC$ stacking structure, it is energetically favourable for a Type 2 surface to undergo a reconstruction in which the second layer switches its pairing partner from the third to the first layer. The uppermost two layers are then paired and the third layer is left un-paired \cite{Lee2023}, as shown in Fig. 13(a). At the edge of a Type 2 surface layer that undergoes this reconstruction, the terrace below the edge is already the energetically preferred Type 1 surface and the pair-switching reconstruction does not propagate into it. Therefore the reconstruction is bounded by a domain wall that is buried exactly underneath the step-edge. Thus two adjacent terraces can both exhibit $\mathbf{T}_{A}$ stacking and a Type 1 spectrum, and the existence of a domain wall buried below the step-edge confounds the estimation of in-plane displacement that would be made by comparing the SD lattice registry between the terraces on either side. The inference from the results of many cleaves \cite{Butler2020,Wu2022} that the formation of the Type 1 surface might be the energetically favored is, in hindsight, consistent with the occurrence of such reconstructions.

\begin{figure}
\centering
\includegraphics[scale=1]{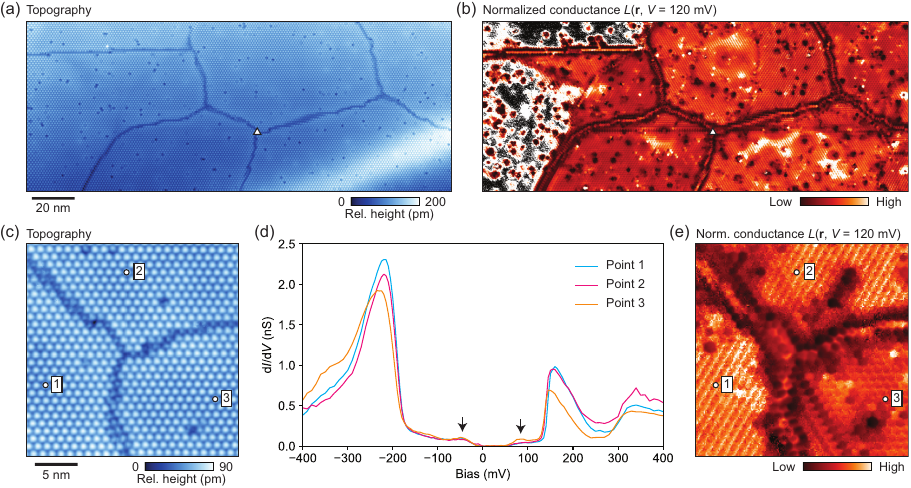}
\caption{\label{fig:14}
(a) STM topography at a Type 1 cleaved surface of 1\textit{T}-TaS$_{2}$ ($T$ = 1.5 K \cite{Hanaguri2006}, $V$ = -500~mV, $I$ = 500~pA). Samples were synthesized and prepared as described previously \cite{Butler2020,Tani1981} (b) Normalized conductance image $L(\mathbf{r}, V) = [\frac{\mathrm{d}I}{\mathrm{d}V}(\mathbf{r},V)]/[I(\mathbf{r}, V)/V]$ in the same field-of-view as (a), acquired within the putative Type 1 band gap at $V$ = 120~mV ($V_{\mathrm{mod.}}$ = 10~mV). The noise in the regions at the left-hand-side occurs due to the nearly-zero current drawn within the ideal Type 1 band gap. The rest of the image shows various textures of `in-gap' DOS, delineated by several buried domain walls that are not observable in the topography. (c) Topography image around the node joining three domain walls, marked with a white triangle in (a). (d) Tunnelling conductance curves acquired at three points as shown in (c). Features inside the expected Type 1 gap, attributed to a buried un-paired SD layer \cite{Yang2024}, are marked with black arrows. Like the Type 2 surface, the band gap is 50$\sim$60~meV. (e) Normalized conductance image in the same field-of-view as (c), at $V$ = 120~mV. The stripey texture in two of the domains may be related to the in-plane displacement of the buried un-paired layer with respect to the overlaid SD bilayer. The differing texture in the remaining domain suggests a further richness of the near-surface stacking degrees of freedom that is yet to be characterized.}
\end{figure}

While the pairing of layers can be reconfigured in such a way, the number of un-paired layers must be constant, so after a reconstruction a buried layer must be left un-paired. Lee \textit{et al.} also showed that, as might be expected intuitively, this buried un-paired layer has an electronic structure somewhat alike to that of the Type 2 surface, i.e. with a reduced gap \cite{Lee2023}. This is shown in Fig. 13(b). Because the states at the edge of the smaller Type 2 gap lie at energies within the larger gap where the overlaid bilayer is `transparent' to tunnelling, the buried un-paired layers are in principle accessible to observation. Yang \textit{et al.} subsequently observed cleaved surfaces with multiple regions in which the tunnelling spectra broadly resemble the Type 1 spectrum, but can be distinguished by the details of states lying within the Type 1 gap \cite{Yang2024}.

In Fig. 14 we present additional observations that mesh well with the explanations advanced in \cite{Lee2023} and \cite{Yang2024}. Figure 14(a) shows STM topography at a Type 1 surface with multiple domain walls, and Fig. 14(b) shows the simultaneously acquired normalized conductance, $L$ (approximately proportional to the DOS), at $V$ = 120~meV, within the Type 1 gap (recall Fig. 8). In the conductance image, buried domain walls can be seen that do not appear in the topography image. Domains at the left-hand-side show either zero conductance or large noise (an artifact of normalization when the tunnelling current is near zero), indicating zero DOS expected for the Type 1 gap. The other domains show some `in-gap' DOS. Figure 14(c) shows a zoom-in view of the intersection between three such domains and Fig. 14(d) shows representative tunnelling spectra acquired in each domain. Each spectrum shows in-gap states (marked with black arrows) separated by a gap of $\sim$50~meV, reminiscent of the Type 2 gap. As the bilayer terminating the Type 1 surface is expected to be transparent to tunnelling at this energy, we can interpret the in-gap states as belonging to a buried un-paired layer. Interestingly, as shown in Fig. 14(e), many of the domains where these in-gap states can be seen show a texture in conductance images with a reduced symmetry as compared to a single CCDW layer, i.e. stripes in Fig. 14(e). This reduced symmetry may be associated with offset stacking between the buried layer and the overlaid Type 1 bilayer.

\begin{figure}
\centering
\includegraphics[scale=1]{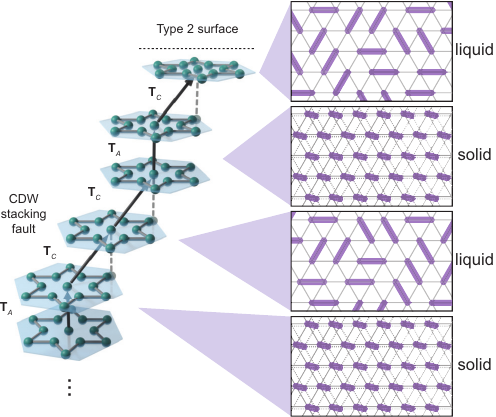}
\caption{\label{fig:15}
Illustration of a possible scenario for isolated QSLs to occur in bilayer-stacked 1$T$-TaS$_{2}$. Un-paired layers, whether at the surface or at stacking faults, can be thought of as two-dimensional Mott insulators and therefore might host a QSL phase \cite{Li2022,ManasValero2021}. On the other hand, the regular bilayer stacking of the bulk should be thought of as a layered spin solid phase.}
\end{figure}

Overall, these observations strongly suggest that un-paired CCDW layers can persist both at the (Type 2) surface and in buried layers in the near-surface region. It is also reasonable to expect that the deep bulk hosts stacking faults of the $AC$ stacking pattern that leave a sparse distribution of un-paired layers. Indeed Monte Carlo simulations representing the quenching of the system into the low-temperature phase support this expectation \cite{Lee2019}. Because there is a strong argument that the mechanism that guarantees insulating behaviour in an un-paired CCDW layer is the Mott mechanism, it is then possible that isolated QSLs may be realized at Type 2 surfaces or stacking faults even if most of the bulk is not believed to be a simple Mott insulator. This possibility has been addressed for the Type 2 surface by Li \textit{et al.}, \cite{Li2022}, and for buried un-paired layers by Ma\~{n}as-Valero \textit{et al.}, who present combined $\mu$-SR and specific heat measurements consistent with several distinct QSL phases residing in sparsely distributed host layers throughout the bulk \cite{ManasValero2021}. A synthesis of these possibilities is sketched in Fig. 15.

%
%
%

\section{Electronic structure of monolayer and few-layer films}

While the nature of the ground state of bulk 1$T$-Ta$X_{2}$ is complicated by inter-layer stacking, these complications can be circumvented, or better understood, in investigations of monolayer or few-layer thin films. There are both top-down an bottom-up ways of producing single- or few-layer 1$T$-Ta$X_{2}$. Top-down methods include the exfoliation of few-layer samples \cite{Tian2024} and, for 1$T$-TaS$_{2}$, potentially also the annealing of bulk crystals up to 600$\sim$700 K which drives a partial polytype transition resulting in a layer-wise mixture of 1$H$- and 1$T$-TaS$_{2}$ \cite{Sung2022}. This might leave a surface layer of the 1$T$ polytype atop 1$H$ layers, but if not, subsequent cleavage might reveal isolated 1$T$ layers. The advantage of these top-down techniques is that they potentially yield relatively large-scale and very clean samples. An additional top-down method is to start with 2$H$-TaS$_{2}$ and locally drive the surface layer through a polytype transition using laser heating, which creates small islands of the 1$T$ phase \cite{Ravnik2021}. Monolayer and few-layer samples can be produced in a bottom-up way using molecular beam epitaxy (MBE) and chemical vapour deposition (CVD) \cite{Lin2018,Lin2020,Vano2021,Ruan2021,Chen2022,Chen2024,Chen2025}. These techniques have been used to prepare samples on both highly oriented pyrolytic graphite (HOPG) and on bilayer graphene on SiC, which both provide a bottom electrode necessary for tunnelling measurements. The graphene layer (or graphite) underlying the TMD film has a Dirac point slightly removed from $E_{\mathrm{F}}$ due to the effect of doping, but still has a very small DOS at low energies and a low carrier concentration. It is then likely that charge transfer and other effects on the transition metal dichalcogenide film are negligible. The energy scale of coupling between the graphene and a TMD layer is likely on the order of only $\sim$meV. While this is relevant for low-energy phenomena \cite{Naritsuka2025}, it can be ignored when discussing the energy gap and other broad features of the electronic structure of 1$T$-Ta$X_{2}$ that we discuss below, which are typically on the $\sim$100~meV scale. In this section we summarize recent STM results on 1$T$-Ta$X_{2}$ and the comparable material 1$T$-NbSe$_{2}$ obtained $via$ bottom-up methods.

\begin{figure}
\centering
\includegraphics[scale=1]{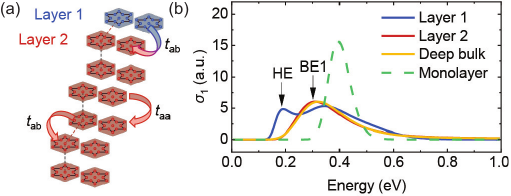}
\caption{\label{fig:16}
(a) Schematic of a seven-layer slab model, featuring a Type 2 surface atop three dimerized bilayers, used in DMFT calculations. (b) The predicted optical conductivity curves (closely related to DOS spectra) for layers near the slab's surface and in the bulk, as compared to that for an isolated monolayer. Although both the Type 2 surface layer (`Layer 1') and the monolayer are unpaired, they exhibit significantly different DOS curves. Reproduced with permission from \cite{Lin2025}. Copyright of the American Physical Society.}
\end{figure}

For 1$T$-TaSe$_{2}$ and 1$T$-TaS$_{2}$ monolayers on a graphene substrate, tunnelling spectra show insulating gaps of about $\sim$100 and $\sim$300~meV respectively \cite{Lin2020,Vano2021,Chen2020}. In the case of 1$T$-TaSe$_{2}$, the gap progressively closes and a metallic state is quickly recovered as the number of layers increases from one to three \cite{Chen2020}. The stacking vectors between layers were seen to belong to the set $\mathbf{T}_{C}$, but varied in in-plane direction, differing with what would be expected for bulk 1$T$-TaSe$_{2}$. Interestingly, for a 1$T$-TaS$_{2}$ monolayer the gap somewhat resembles that of the layer-wise dimerized Type 1 surface \cite{Lin2020,Vano2021}, and is much larger than that for the Type 2 surface, with which it would na\"{i}vely be expected to be more similar. A possible explanation for this apparent discrepancy is found in recent DMFT results reported by Lin \textit{et al.}, for the electronic structure of paired and unpaired near-surface layers as well as an isolated monolayer \cite{Lin2025} [see Fig. 16]. The reduction of the gap size for the Type 2 surface was specifically attributed to significant inter-layer hopping across the $\mathbf{T}_{C}$ stacked interface between the uppermost and underlying layers, an effect which is absent for the isolated monolayer.

An important aspect of the electronic structure of monolayer 1$T$-Ta$X_{2}$ and 1$T$-NbSe$_{2}$ that cannot be overlooked is the presence of an additional gap (as well as the insulating gap around $E_{\mathrm{F}}$) that appears in tunnelling conductance curves at around +500~meV. Below and above this gap are a peak and a shoulder in the unoccupied states that are each identified as UHBs and are labelled as UHB$_{1}$ and UHB$_{2}$ respectively \cite{Lin2020,Vano2021,Chen2020,Nakata2016,Liu2021a,Liu2021b}. This apparent occurrence of multiple UHBs is accompanied by intriguing behaviour in the spatial distribution of unoccupied states. Figure 17(b) presents a tunnelling spectrum for monolayer 1$T$-TaSe$_{2}$, as well as corresponding tunnelling conductance maps of the salient spectral features. According to Chen \textit{et al.}, conductance mapping shows that in the occupied states at about $-150$~meV, namely the LHB, closely resembles that measured at the surface of bulk samples. However, at the energies of spectral features identified as UHB$_{1}$ and UHB$_{2}$, the maxima of conductance are not located at the SD centres but instead form a pattern around the SD cluster periphery, indicating a more complicated orbital texture. Such orbital textures are seen in monolayers but not in bi- or trilayer samples, and their origin remains mysterious. In 1$T$-NbSe$_{2}$, similar distinct patterns have been observed and are attributed to a temperature-induced structural distortion \cite{Liu2021c}. A slightly different intra-unit-cell structure for the CDW of the monolayer system has also been suggested \cite{Park2023b}. The effect of surface-adsorbed (non-magnetic) impurities, which could shed light on the nature of the insulating state, has been investigated for the monolayer \cite{Chen2022} as well as the bulk \cite{Lee2021}, but a complete picture remains elusive. For monolayer 1$T$-NbSe$_{2}$, there is an ongoing debate as to whether it should more correctly be classified as a Mott insulator \cite{Nakata2016,Liu2021a} or a charge-transfer insulator \cite{Liu2021b} (and a similar question persists for the insulating behaviour at the surface of bulk 1$T$-TaSe$_{2}$ \cite{Sayers2023}).

\begin{figure}
\centering
\includegraphics[scale=1]{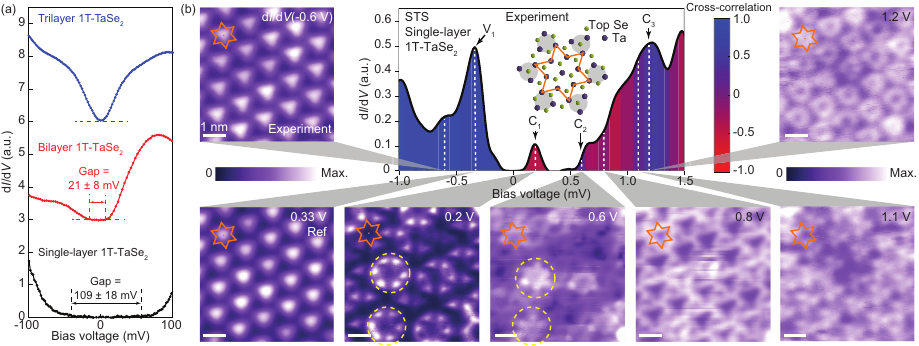}
\caption{\label{fig:17}
(a) Thickness dependence of tunnelling spectra for 1$T$-TaSe$_{2}$ on graphene/SiC. The insulating gap rapidly closes as the number of layers increases. (b) Large energy range tunnelling spectrum for monolayer 1$T$-TaSe$_{2}$, and tunnelling conductance maps at selected energies. The patterns exhibit an inversion of intensity between the occupied and unoccupied states, a feature which is not captured using a simple DFT-based model. Adapted with permission from \cite{Chen2020}. Copyright of Springer Nature.}
\end{figure}

While 1$T$ monolayers of Ta$X_{2}$ or NbSe$_{2}$ are only negligibly influenced by an underlying HOPG or graphene substrate, the case is very different for a 1$T$ monolayer on top of a 1$H$ monolayer on graphene \cite{Vano2021,Ruan2021}, or for 1$T$-NbSe$_{2}$ formed at the surface of 2$H$-NbSe$_{2}$ by a local electric field pulse \cite{Liu2021b}. In these cases the 1$T$ monolayer shows metallic behaviour. One explanation is the larger DOS and carrier density as compared to graphene or graphite, which allows an enhanced Kondo coupling (see below) between the uppermost and underlying layers, that serves as an enhanced effective hopping between adjacent SD cluster in the overlayer. Another possible explanation is charge transfer from the 1$T$ to the 1$H$ layer that significantly depletes $U$, leading to collapse of the Mott state \cite{Bae2025}. However, the detailed mechanism driving metallicity in this system is yet to be established.

\subsection{Kondo physics}

In metallic system with sparsely distributed magnetic impurities, the impurity spin can interact with the spins of conduction electrons \textit{via} exchange coupling, leading to the formation of a many-body singlet state in which the impurity spin is collectively screened by surrounding electrons [see Fig. 18(a)]. This singlet state manifests in the single-particle excitation spectrum as a narrow DOS peak near $E_{\mathrm{F}}$, referred to as a Kondo resonance peak. STM has been widely used to investigate the spatial properties of Kondo resonance states. Examples include 3$d$ transition metal and rare earth metal atoms \cite{Li1998,Madhavan1998}, and metal phthalocyanine molecules with magnetic ions \cite{Zhao2005,Franke2011}, adsorbed on non-magnetic metal substrates.

Recently, Kondo resonance states have been reported in 1$T$-TaS$_{2}$/1$H$-TaS$_{2}$ hetero-structures, attracting attention as a new platform for the investigation of Kondo physics \cite{Ruan2021,Vano2021,Wan2023}. Va\v{n}o \textit{et al.} observed a Kondo resonance near $E_{\mathrm{F}}$ in the metallic spectrum of the 1$T$ overlayer, and observed in conductance maps that the Kondo peak is concentrated at the SD cluster centres. The shape and temperature-dependent evolution of the of the spectrum follows the Fano-Kondo function with an estimated Kondo temperature of $T_{\mathrm{K}}$ = 30~K. This represents strong evidence that SD clusters each host a single localized spin.

As the concentration of magnetic impurities increases, the magnetic interactions between localized spins can become significant. A particularly interesting magnetic interaction in this context is the relatively long-ranged Ruderman-Kittel-Kasuya-Yosida (RKKY) interaction. While the Kondo effect screens localized spins and suppresses magnetism, the RKKY interaction tends to stabilize magnetic order, and the two interactions are in competition. Theoretical studies on two-impurity Kondo systems predict that the sign of the RKKY interaction can lead to changes in the width, or the splitting, of the Kondo peak. Furthermore, when magnetic impurities form a lattice, they can give rise to a Kondo lattice, which serves as a platform for exotic physics including quantum criticality, non-Fermi liquid behaviour, heavy fermion states, and unconventional superconductivity.

\begin{figure}
\centering
\includegraphics[scale=1]{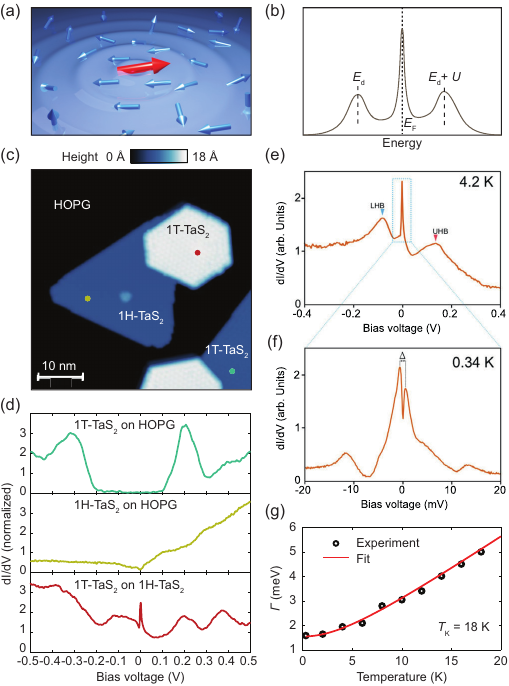}
\caption{\label{fig:18}
(a) Illustration of Kondo screening of a single magnetic impurity (red) in the Fermi sea carrying itinerant spins (blue). (b) Typical Kondo resonance peak (around $E_{\mathrm{F}}$) and LHB and UHB peaks (at $E_{\mathrm{d}}$ and $E_{\mathrm{d}} + U$) in the local DOS. (c) STM topography of few-layer TaS$_{2}$ grown on HOPG. (d) Tunnelling spectroscopy of monolayer 1$T$-TaSe$_{2}$ on HOPG, 1$H$-TaSe$_{2}$ on HOPG, and a 1$T$/1$H$ hetero-bilayer. (e) Tunnelling conductance on acquired on a 1$T$/1$H$ hetero-structure at $T$ = 4.2~K. The position of the LHB and UHB are indicated. (f) Low-bias conductance at $T$ = 0.34~K, showing a splitting of the Kondo peak. (g) Temperature dependence of Kondo resonance width (black circles). The red curve corresponds to fitting using the Fano-Kondo function and the resulting temperature is $T_{\mathrm{K}}$ = 18~K. (c), (d) and (g) reproduced from \cite{Vano2021}. Copyright of Springer Nature. (e) and (f) reproduced from \cite{Wan2023} under CC BY 4.0 license.}
\end{figure}

Chen \textit{et al.}, propose that such a Kondo lattice is realized in 1$T$/1$H$ hetero-structures. In these systems, the SD lattice spacing is about 12.1~\AA, a distance over which the RKKY interaction is expected to be effective. Theoretically, a heavy fermion state could emerge in the 1$T$/1$H$ system. To further investigate the coherence of the electronic state in the system, Ayani \textit{et al.} conducted low-temperature tunnelling spectroscopy near $E_{\mathrm{F}}$ and examined their magnetic field dependence \cite{Ayani2022}. Their results show peak splitting, raising the question of whether a Kondo lattice is formed or if itinerant magnetic order has developed.

While clear experimental evidence of Kondo coherence remains elusive, future STM tunnelling spectroscopy investigations on isolated thin films and hetero-structures may provide more concrete evidence for it.

\section{Evidence for QSL behaviour from STM measurements}

\subsection{Possible spinon-chargon bound states at the surface of bulk 1$T$-TaS$_{2}$}

\begin{figure}
\centering
\includegraphics[scale=1]{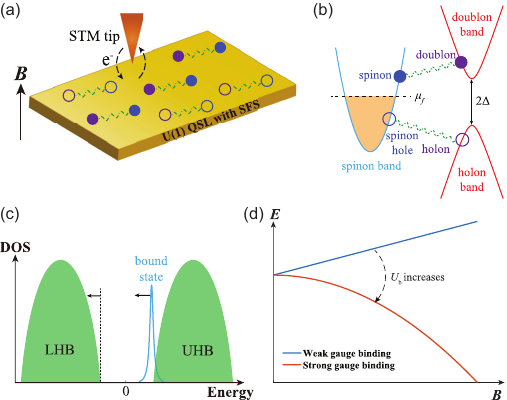}
\caption{\label{fig:19}
(a) Electrons injected into a QSL from an STM tip dissociate into spinons and chargons (doublons and holons). The filled blue and purple dots denote spinons and doublons, while the empty dots denote spinon holes and holons, respectively. The wavy lines indicate an attractive interaction due to fluctuations of a U(1) gauge field. (b) Spinons might reside in a parabolic band that can generally span a part, or the whole, of the charge gap. (c) DOS spectrum of the Mott insulator with a spinon-doublon bound state at the edge of the UHB. (d) Theoretically predicted magnetic field dependence of the bound state energy in the weak and strong gauge binding regimes. Reproduced with permission from \cite{He2023a}. Copyright of the American Physical Society.}
\end{figure}

\begin{figure}
\centering
\includegraphics[scale=1]{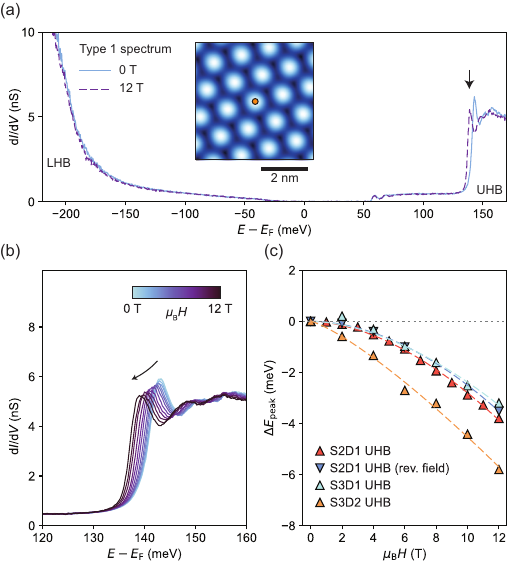}
\caption{\label{fig:20}
(a) Field-dependent tunnelling conductance curves acquired at a Type 1 cleaved surface of bulk 1$T$-TaS$_{2}$. A narrow peak appears at the edge of the feature often identified as the UHB, and shifts to a lower energy under external magnetic field. Interestingly, the in-gap DOS described above also appears here. (b) Zoom-in view of the field dependence of the narrow peak and (c) quantitative field dependence obtained using a phenomenological fit for the peaks found at multiple different locations and samples. The field dependence is approximately quadratic, against expectation for a Zeeman shift, and the shift is towards lower energy, contrary to expectation for the Landau level of an electron-like band. The behaviour is however compatible with that described for a spinon-holon bound state as described in \cite{He2023a} and illustrated in Fig. 19(d). Reproduced with permission from \cite{Butler2023}. Copyright of the American Physical Society.}
\end{figure}

As introduced in Section 2.2, although a QSL does not have an order parameter that can be measured with any local microscopic probe, recent theoretical investigations of a particular type of QSL have proposed that spinons might be amenable to local experimental detection. Although the type of QSL thought most likely to emerge in the triangular Heisenberg antiferromagnet is a Dirac QSL that has a gapless spinon spectrum but no spinon Fermi surface, alternative models start by specifying the existence of a spinon Fermi surface, and this can have an imprint on the electronic DOS spectrum that is measurable using tunnelling spectroscopy \cite{Tang2013,He2023a,He2023b}. This imprint might arise through fluctuations of an emergent gauge field that couples to both spinons residing in a spinon band and chargons near the edges of the charge gap (i.e. in the UHB or LHB), as depicted in Fig. 19. The fluctuations result in an attractive interaction that serves to bind spinons and chargons into electrons, and this can form bound states that are separate from the ordinary UHB or LHB continuum. Moreover, the resulting bound states were predicted by He and Lee \cite{He2023a} to have a magnetic field-dependent binding energy, shifting to lower energies with increasing external field, and with a quadratic dependence for sufficiently strong `gauge binding', as illustrated in Fig. 19(d).

As mentioned in Section 4.2, some tunnelling conductance measurements at the surfaces of bulk 1$T$-TaS$_{2}$ have detected unusual, narrow peaks near the UHB edge \cite{Butler2021}. Interestingly, recent magnetic field dependent tunnelling spectroscopy measurements have shown an approximately quadratic field-dependence of the peak energy, shown in Fig. 20, which is consistent with predictions of He and Lee \cite{Butler2023,He2023a}.

Paradoxically, the predicted field-dependence of the peak energy was observed at the Type 1 surface, which should be less likely to host a QSL due to the implied intra-bilayer dimerization. Some sample-dependence of the effect was observed -- Both the presence and absence of the effect were seen at various Type 1 surfaces. The corresponding peak in the spectrum for the Type 2 surface was observed, but magnetic field-dependence was not. However, the reproducibility of this absence was not established. An effort to do so is still needed.

\subsection{Kondo resonance of magnetic impurities in a spinon fluid}

\begin{figure}
\centering
\includegraphics[scale=1]{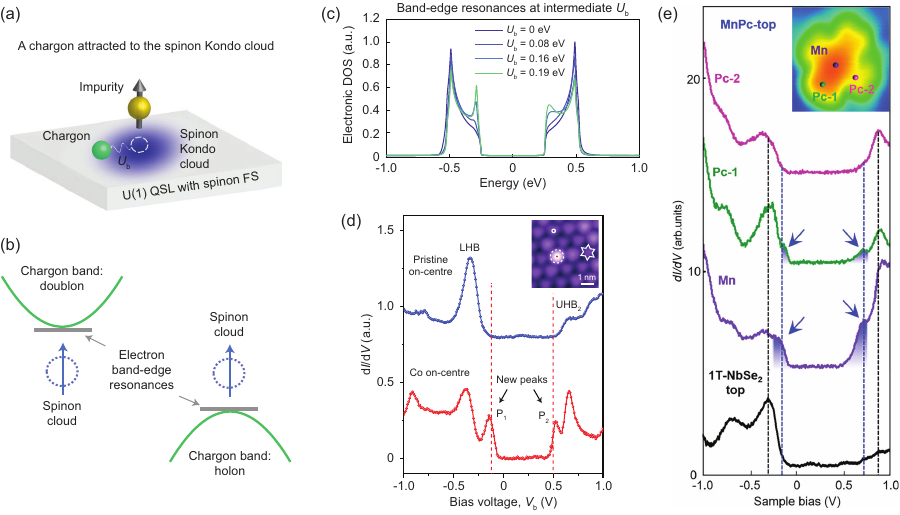}
\caption{\label{fig:21}
(a) Schematic of Kondo screening of a single magnetic impurity in a spinon Fermi sea. (b) Schematic of resonance states at the edges of the chargon (doublon and holon) bands. (c) The calculated impurity electronic DOS spectrum for different binding interactions $U_{\mathrm{b}}$. (d) Tunnelling spectroscopy of pristine monolayer 1$T$-TaSe$_{2}$ and of an adsorbed Co atom. Inset: STM topography showing the Co atom atop a SD cluster. Adapted from \cite{Chen2022}. (e) Tunnelling spectroscopy of pristine monolayer 1$T$-NbSe$_{2}$ and of an adsorbed MnPc molecule. The different coloured curves correspond to different locations on the molecule, as marked in the inset. Reproduced from \cite{Zhang2024} under CC BY 4.0 license.}
\end{figure}

A QSL with a spinon Fermi surface is insulating in the charge channel but metallic in the spin channel. The behaviour for a single magnetic impurity in a two-dimensional QSL has been investigated theoretically, and it was found that the Kondo effect involving spinons can occur through a mechanism similar to that in a Fermi liquid, despite the fact that it is a charge insulator \cite{Ribeiro2011}. Experimental signs of this spinon Kondo effect would provide evidence for a QSL with a spinon Fermi surface \cite{He2022}.

Chen \textit{et al.} deposited magnetic cobalt atoms on 1$T$-TaSe$_{2}$ monolayer samples and investigated the impurity spectra using tunnelling spectroscopy \cite{Chen2022}. Indeed, resonance peaks were observed simultaneously at the UHB and LHB edges when a Co atom was located on top of a SD centre \cite{Chen2022}. These resonance peaks disappeared when the Co atom was located away from the SD centre, and were not observed for non-magnetic impurities such as K or Au. It was argued that this phenomenon is due to the formation around magnetic impurities of a Kondo cloud due to Kondo screening by itinerant spinons. Subsequently the spinon Kondo cloud affects chargons $via$ fluctuations of the gauge field [see Fig. 21(a)], in a manner similar to that described above, causing resonances to appear at the edge of the UHB and LHB. Such resonance peaks do not appear in the absence of gauge binding and their intensity depends on its strength \cite{He2022}.

Zhang \textit{et al.} reported similar magnetic impurity-induced states at the UHB and LHB edges in the impurity spectrum of manganese phthalocyanine (MnPc) molecules (with spin $S$ = $\frac{3}{2}$) on monolayer 1$T$-NbSe$_{2}$ \cite{Zhang2024}. When a MnPc molecule rests on top of the centre of a SD cluster, a pair of peaks appears at the edges of the features identified as the LHB and UHB$_{2}$. It is spatially concentrated at the Mn atom at the centre of the Pc molecule and greatly diminished in the molecule's outer lobes. In contrast, no such pair of peaks is observed for the non-magnetic ZnPc molecule. This contrast supports the scenario of a spinon Kondo effect arising from a magnetic impurity placed in a QSL with a spinon Fermi surface.

The possible observation of the spinon Kondo effect in 1$T$-TaSe$_{2}$ and 1$T$-NbSe$_{2}$ suggests that the system hosts a gapless QSL. In gapped spin liquids, such as the $Z_{2}$ spin liquid, the spinon Kondo effect is not expected to occur. Although a Dirac spin liquid is also gapless, its stability requires the presence of an external gauge field flux \cite{Zhou2017,Hu2019}. The fact that the spinon Kondo effect is observed under zero external magnetic field makes it unlikely that the system is a Dirac spin liquid. Therefore, the experimental results point to the realization of a QSL with a spinon Fermi surface \cite{He2022}.

\subsection{Signatures of QSL behaviour in monolayer 1$T$-TaSe$_{2}$}

Ruan \textit{et al.}, reported a more direct manifestation of the characteristics of a spinon Fermi surface in monolayer 1$T$-TaSe$_{2}$ \cite{Ruan2021}. At the energies of the LHB and both UHBs, an incommensurate super-modulation (ICS) overlaid on the SD lattice. A direct connection was inferred between this ICS and the wave vector of the spinon Fermi surface as calculated using a triangular lattice $t$-$J$ model (a derivative of the aforementioned Hubbard model). Furthermore, energy-resolved observation of the Fourier components of the ICS indicated a partially gapped spinon Fermi surface. The observed wavevectors correspond to harmonics of the spinon Fermi surface wavevectors suggesting a spinon pair density wave. The peak intensity of the modulation decreased with increasing temperature and was not observed above 80~K. This suggests an origin separate from that of the underlying CDW, which disappears at $\sim$300~K. In addition to the ICS, a commensurate $\sqrt{3} \times \sqrt{3}$ super-modulation (CS) imposed on the SD lattice was observed in conductance maps. While the origin of this feature is unclear, short-ranged magnetic order may be involved. The fact that the $120^{\circ}$ antiferromagnetic ordered state [recall Fig. 2(d)] is adjacent to the inferred U(1) spin liquid in the magnetic phase diagram is suggestive.

\begin{figure}
\centering
\includegraphics[scale=1]{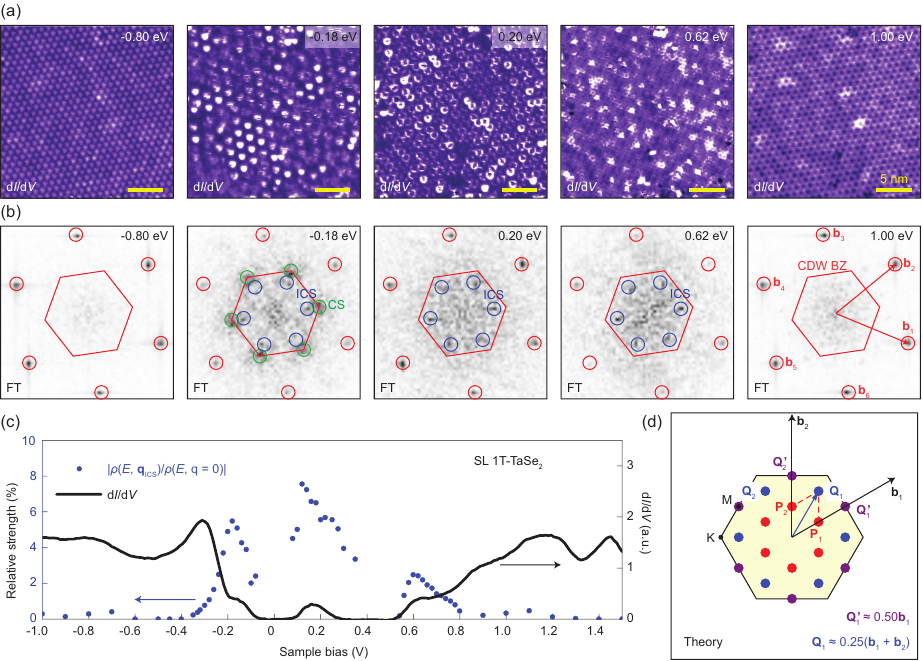}
\caption{\label{fig:22}
(a) Tunnelling conductance maps at selected energies acquired on monolayer 1$T$-TaSe$_{2}$ and (b) their respective Fourier transforms (FT). Red circles represent reciprocal lattice vectors of the SD CDW. Red hexagons represent the CDW BZ. The peaks circled in blue in the FT images reflect incommensurate super-modulations (ICS), and those in green reflect commensurate super-modulations (CS). (c) Energy-dependence of the amplitude of the Fourier peaks for the ICS (blue circles), and the conductance, d$I$/d$V$ (black curve). The ICS amplitude shows enhancement at the Hubbard band edges. (d) The theoretically predicted wavevectors for the spinon Fermi surface instability $\mathbf{P}_{i}$ (red), and the higher harmonics $\mathbf{Q}_{i}$ and $\mathbf{Q}_{i}^{'}$ ($1 \leq i \leq 6$) in the CDW BZ. Reproduced with permission from \cite{Ruan2021}. Copyright of Springer Nature.}
\end{figure}

Zhang \textit{et al.} performed STM conductance mapping on 1$T$-NbSe$_{2}$ and obtained results similar to those for 1$T$-TaSe$_{2}$ \cite{Zhang2024}. In Fourier transformed conductance images, the CDW wavevectors dominate at energies far from the UHB and LHB edges. However, near the edges of LHB, UHB$_{1}$ and UHB$_{2}$, an additional charge ordering seems to emerge. This ordering is incommensurate, with a periodicity slightly larger than $\sqrt{3}$ times the CDW periodicity, and with lattice vectors rotated by $30^{\circ}$ with respect to those of the CDW. These findings closely resemble those reported for 1$T$-TaSe$_{2}$ in terms of energy range, characteristic wavevectors, and incommensurability with the CDW. Importantly, the quasiparticle interference measurements show that the incommensurate scattering vectors observed in 1$T$-NbSe$_{2}$ and 1$T$-TaSe$_{2}$ have almost identical lengths. While the origin of this coincidence in the characteristic wavevector of the underlying mechanism, such as the putative spinon Fermi surface size, is not yet fully understood, it indicates that it is a robust feature of these systems. Furthermore, the close similarity between Nb- and Ta-based compounds highlights that the phenomenon is resilient to changes in the transition-metal species. This robustness provides important context for material exploration, as it suggests that the salient phenomenon emerges in a way that it somewhat independent of material details and can be realized across a broad class of 1$T$-transition metal dichalcogenides.

\subsection{Commensurate and incommensurate DOS modulations in monolayer 1$T$-TaSe$_{2}$}

\begin{figure}
\centering
\includegraphics[scale=1]{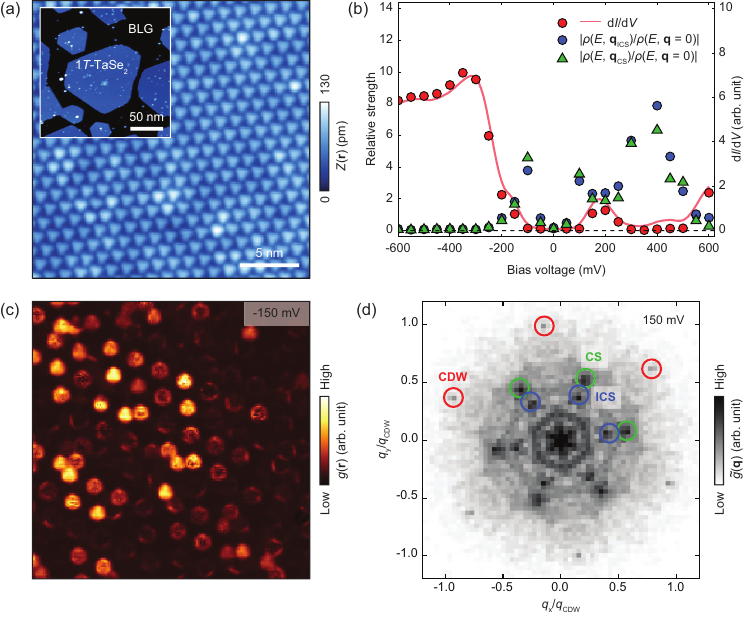}
\caption{\label{fig:23}
(a) A typical topographic image $Z(\mathbf{r})$ of MBE-grown monolayer 1$T$-TaSe$_2$ on bilayer graphene (BLG), showing the SD CDW. The image was taken with a feedback set-point current $I_\mathrm{set}$ = 100~pA at a sample bias voltage $V_\mathrm{set}$ = -600~mV. Inset: Topography taken in a larger area using the same set-point and bias.
(b) Bias dependence of Fourier peak intensities at the ICS wavevector $\rho(E,\mathbf{q}_\mathrm{ICS})$ (blue circles), and the CS wavevector $\rho(E,\mathbf{q}_\mathrm{CS})$ (blue circles). These intensities are divided by the Fourier amplitude at $q$ = 0, $\rho$($E$,$q$=0). The reference d$I$/d$V$ conductance spectrum taken at the same time are plotted in red circles. Light red line is the average d$I$/d$V$ spectrum taken at the top of $\sim$80 SD clusters.
(c) The conductance map $g$($\mathbf{r}$, $V$=-150~mV). The measurement conditions were $I_\mathrm{set}$ = 100~pA, $V_\mathrm{set}$ = -600~mV, and the bias modulation amplitude was $V_\mathrm{mod}$ = 20~mV.
(d) Fourier transformed $\tilde{g}$($\mathbf{q}$, $V$=+150~mV). Red, green, and blue circles represent the locations of CDW, CS, and ICS, respectively.}
\end{figure}

Here we complement the results reviewed above with additional observations acquired on 1$T$-TaSe$_2$ grown on a graphene/4$H$-SiC substrate using MBE. Measurements were performed at $T \approx$ 4.2~K using an STM instrument described previously \cite{Machida2017}. The measurements focused on regions within 1$T$-TaSe$_2$ islands where the CDW was found to be uniform over a large field-of-view [see Fig. 23(a)]. The tunnelling conductance spectrum acquired at the centre of an SD cluster shows an insulating gap of approximately $\pm$100~mV as shown in Fig. 23(b), consistent with previous reports \cite{Chen2020, Lin2020}. Figure 23(c) shows the differential conductance $g$($\mathbf{r}$,$E$=eV) $\equiv$ d{$I$($\mathbf{r}$,$V$)}/d{$V$} map near the Hubbard band edge, which exhibits an apparent $\sqrt3\times\sqrt3$ CS. Fourier analysis reveals not only this CS pattern but also an ICS with $\mathbf{q}_\mathrm{ICS}\sim0.245(\mathbf{b}_1 + \mathbf{b}_2)$, where $\mathbf{b}_i$ is the unit vector for the CDW [Fig. 23(d)]. Our value of $\mathbf{q}_\mathrm{ICS}$ is very close to the value $\mathbf{q}_\mathrm{ICS}\sim0.241(\mathbf{b}_1 + \mathbf{b}_2)$ reported Ruan \textit{et al.} \cite{Ruan2021}. Similar to their results, the ICS peak intensity in our measurements exhibits an enhancement near the Hubbard band edges [see Fig.23(b)].

\begin{figure}
\centering
\includegraphics[scale=1]{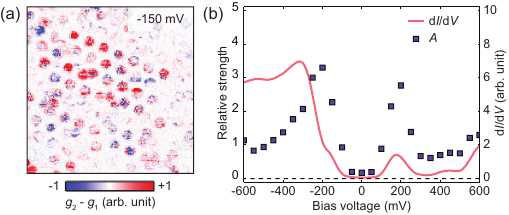}
\caption{\label{fig:24}
(a) A difference map obtained between two $g$ maps taken one after another in the same field-of-view as for Fig. 23(b). (b) Energy dependence of the total variation of amplitude $A$. The light red curve is the average tunnelling spectrum.}
\end{figure}

While the observations of the CS modulation reported by Ruan \textit{et al.} were limited to negative energies, we observed a pronounced CS intensity at both positive and negative energies [Fig. 23(b)]. Moreover, the intensities of the ICS and CS seem to be closely related. The origin of the CS modulation is an open question. Ruan \textit{et al.} suggested that it may arise from short-range antiferromagnetic fluctuations, given that the QSL is near the 120$^{\circ}$ antiferromagnetic phase in the phase diagram \cite{Ruan2021}. However, in this case, the three magnetic sub-lattices are equivalent, so na\"{i}vely, one would not expect a $\sqrt{3} \times \sqrt{3}$ superstructure to appear in the conductance maps. Lattice strain in the atomically thin film samples is one possibility that could render the sub-lattices inequivalent. Other possible scenario include surface adsorption or intercalation of excess atoms or molecules \cite{Lee2021}, or the realization of a Wigner crystal-like electron distribution in a fractionally filled lattice \cite{Huang2021}. In any case, elucidating the origin of the CS periodicity that coexists with ICS remains a future challenge.

Notably, the local DOS exhibited variations upon repeated measurements. Figure 24(a) shows a $g(\mathrm{r},V)$ difference map obtained from two $g$ maps measured sequentially in the same area, with red indicating regions of increased intensity and blue indicating decreased intensity. Changes occur on a SD unit-by-unit basis. To quantitatively evaluate these variations, we define a total variation amplitude, $A(V) = \Sigma_{(x,y)}|g_2(x,y,V) - g_1(x,y,V)|$, as the sum of absolute value of pixel-by-pixel intensity changes within the measured area. The energy dependence of this variation is plotted in Fig. 23(b). Significant changes are observed near the energies corresponding to the gap edges. These scan-to-scan variations suggest the presence of many nearly degenerate electronic states localized on individual SD clusters. Due to the small energy barriers separating these states, even minor perturbations -- such as the electrostatic potential or electric field induced by the proximity of the STM tip -- may induce transitions between them. This sensitivity reflects the delicate nature of the electronic structure inherent to 1$T$-TaSe$_2$.

\section{Summary and Perspective}

In Section 4 we have seen that STM and other experimental results amount to quite a strong argument that an alternating inter-layer stacking is indeed realized in bulk 1$T$-TaS$_{2}$, and hence if a QSL is realized anywhere, it is most likely at surfaces or stacking faults where the possible inter-layer dimerization is broken, as illustrated in Fig. 15. In light of the realization of the bilayer stacking pattern it is sometimes stated that the bulk is simply a band insulator \cite{Wang2020}. Still, phenomena consistent with strong electronic correlations observed in ultra-fast dynamics \cite{Perfetti2008,Petersen2011,Mann2016,Ligges2018,Zhang2019,Avigo2020}, STM \cite{Bu2019} and other measurements \cite{Dong2023} cannot be dismissed. Also, the argument that a single 1$T$-TaS$_{2}$ layer is a Mott insulator is very strong, and this does have relevance to the bulk properties: It implies a high effective Mottness ratio $U/W$, wherein $U$ should be fairly insensitive to the stacking environment, i.e. it is not diminished by a bilayer stacking configuration. (Observed insulator-metal transitions involving only 1$T$-TaS$_{2}$ layers are probably attributable to a stacking-driven enhancement of $W$ rather than a decrease of $U$.) Then, is it sensible to think that an aggregate of many layers, each of which would be a Mott insulator in isolation, stack up to become some kind of insulator other than a Mott insulator?

In principle the band and Mott insulating mechanisms are not mutually exclusive. A subject of theoretical investigation which is very relevant here is the dimer Hubbard model \cite{Najera2018}, which features two planar lattices hosting the usual in-plane hopping $t$, and also an inter-layer hopping $t_{\perp}$ that connects each site with its counterpart in the other lattice. It has been shown using DMFT that, upon tuning $t_{\perp}$, the dimer Hubbard model can exhibit a continuous cross-over between extremes of Mott insulating and dimerized band insulating nature, and as such the two insulating states can be thought of as coexisting with no meaningful demarcation between them \cite{Fuhrmann2006}. This suggests that it is overly simplistic to think of bulk 1$T$-TaS$_{2}$ as a simple band insulator.

While in the bulk there is a richness of inter-layer stacking degrees of freedom that eludes a full understanding, and in the monolayer there is much greater simplicity (and a greater expectation of finding a QSL), there is strong motivation to investigate an intermediate system, if possible: Isolated bilayer 1$T$-TaS$_{2}$ should be a particular target of experiments because it is the minimal system for investigating the impact of inter-layer interactions. Not only is it apt to be modelled using the dimer Hubbard model, but it may also make the inter-layer hopping parameter $t_{\perp}$ amenable to manipulation (e.g. through electric pulse-induced changes to the stacking) in such a way that the system might be switched between band insulating-like, correlated metal, and Mott-like phases. Very recently Bae and colleagues have predicted further interesting behaviours including flat bands and a Kondo insulating phase, and stacking-mediated transitions between them, that may become available in a 1$T$/1$T$/1$H$ trilayer \cite{Bae2025}.

In Section 6 we have reviewed multiple pieces of evidence that suggest the presence of a spinon Fermi surface both at the surface of bulk 1$T$-TaS$_{2}$ and in monolayer films of 1$T$-Ta$X_{2}$ and 1$T$-NbSe$_{2}$. Remaining questions whose answers are well within reach using present measurement techniques include the following: Can the magnetic-field dependent energy of the narrow resonance at the UHB edge (see Fig. 20) \cite{Butler2023}, consistent with that predicted for the binding energy of a spinon-chargon bound state \cite{He2023a}, be seen at both the Type 1 and Type 2 surfaces of the bulk? Also, can the same narrow resonance be found in monolayer samples, and if so, does it exhibit the predicted magnetic field dependence? 

Aside from a spinon Fermi surface, another possibility that is worth considering is that of a random singlet state \cite{Kimchi2018a,Kimchi2018b}, described in Section 2.2. This state is closely related to quenched disorder that can be characterized using microscopic techniques such as STM, and may also have characteristic time evolution similar to a glassy state \cite{Pal2020}. However, at present we are aware of no explicit predictions for microscopic observables that could signify a random singlet state.

To search for the impacts of a spinon Fermi surface or other potential exotic states on the local DOS, not only tunnelling spectroscopy but also spectroscopic imaging of electronic modulations has provided intriguing evidence \cite{Ruan2021}. However, a clearer characterization of these modulations is needed. Synthesizing and investigating larger, cleaner monolayer samples should allow better characterization of the long-wavelength commensurate and incommensurate orders, and important properties such as their spatial coherence and fluctuations, their magnetic field dependence, and so on. There is also a strong motivation to explore the exfoliation or growth of thin 1$T$-Ta$X_{2}$ and 1$T$-NbSe$_{2}$ films on substrates other that graphite or graphene \cite{Chen2024}. The choice of substrate has been suggested as a way of controlling electronic properties such as the stability of the various CDW phases, and tuning the transitions between them \cite{Zhao2017}. It may also be significant in the search for a QSL phase, as it has recently been suggested that the dielectric properties of the substrate may serve to tune the degree of magnetic frustration \cite{ChenRosnerLado2022}. As previously mentioned, the nature of the insulating state of a material can be elucidated using time-resolved measurements of its dynamics \cite{Hellmann2012}. In the case of few-layer thin films, this could be enabled by newly developed techniques combining femtosecond optical or Terahertz pulses with STM \cite{Cocker2021,Muller2024}, and as well as investigations of monolayer samples, investigations of the stacking-dependent electronic dynamics of few-layer 1$T$-TaS$_{2}$ should yield interesting insights. With respect to the detection of a spinon Fermi surface in particular, it has been suggested that the system's characteristic noise and other properties could be detected through coupling to a qubit-based probe, such as in nitrogen-vacancy (NV) centre which could in principle be extended to scanned NV centre microscopy techniques \cite{Chatterjee2019,Khoo2021,Khoo2022,Lee2023b}.

To conclude, both bulk 1$T$-TaS$_{2}$ and thin film 1$T$ metal dichalcogenides hosting SD lattices are deserving of intensified investigations using state-of-the-art STM. In particular, there are many opportunities that will become available with the improvement and development of novel thin-film fabrication methods, both bottom-up and top-down. Further investigations might gather more concrete evidence for (or against) QSL behaviour, but at minimum will provide powerful insights into two-dimensional strongly correlated materials as both theoretically intriguing and potentially technologically useful systems.

\section*{Acknowledgements}
We thank M Yoshida and Y Iwasa for providing 1$T$-TaS$_{2}$ single crystals, and T Machida for assistance with measurements on monolayer 1$T$-TaSe$_{2}$.
We are grateful to 
Y Kohsaka,
J Lee,
H W Yeom,
M Hirayama,
R Arita,
Y Nomura,
N Nagaosa,
M Civelli,
V Dobrosavljevi\'{c},
M J Rozenberg,
D Mihailovic,
V Va\v{n}o,
W Y He,
P A Lee,
and A Furusaki
for fruitful discussions.

This work was supported in part by JSPS KAKENHI grant number JP19H01855, and by the RIKEN TRIP initiative (Many-body Electron Systems). C.J.B. and M.N. acknowledge support from RIKEN's Programs for Junior Scientists.


\end{document}